\documentclass[preprint,12pt]{elsarticle}
\usepackage{cite}
\usepackage{hyperref}

\usepackage[final]{changes}
\makeatletter
\@namedef{Changes@AuthorColor}{red}
\colorlet{Changes@Color}{red}
\makeatother

\usepackage{amsmath}
\usepackage{amssymb}
\usepackage{dsfont}
\usepackage{multirow}
\usepackage{xfrac}
\usepackage{subcaption}

\usepackage{lineno}

\usepackage{calc}

\usepackage{soul,xcolor}

\usepackage[normalem]{ulem} 
\usepackage{xcolor}\definecolor{RED}{rgb}{1,0,0}\definecolor{BLUE}{rgb}{0,0,1} 
\providecommand{\DIFadd}[1]{{\protect\color{black}#1}} 
\providecommand{\DIFaddbegin}{} 
\providecommand{\DIFaddend}{} 
\providecommand{\DIFdelbegin}{} 
\providecommand{\DIFdelend}{} 
\providecommand{\DIFdel}[1]{} 
\begin{document}

\begin{frontmatter}



\title{U-FNO - an enhanced Fourier neural operator-based deep-learning model for multiphase flow}

\author[inst1]{Gege Wen}
\author[inst2]{Zongyi Li}
\author[inst3]{Kamyar Azizzadenesheli}
\author[inst2]{Anima Anandkumar}
\author[inst1]{Sally M. Benson}

\affiliation[inst1]{organization={Energy Resources Engineering, Stanford University},
            addressline={367 Panama St}, 
            city={Stanford},
            postcode={94305}, 
            state={CA},
            country={USA}}
            
\affiliation[inst2]{organization={Computing and Mathematical Sciences, California Institute of Technology},
            addressline={1200 E. California Blvd., MC 305-16}, 
            city={Pasadena},
            postcode={91125}, 
            state={CA},
            country={USA}}
            
\affiliation[inst3]{organization={Department of Computer Science, Purdue University},
            addressline={305 N University St}, 
            city={West Lafayette},
            postcode={47907}, 
            state={IN},
            country={USA}}


\begin{abstract}
Numerical simulation of multiphase flow in porous media is essential for many geoscience applications. \DIFdelbegin \DIFdel{Data-driven machine learning methods}
\DIFdel{ provide faster alternatives to traditional simulators }
\DIFdel{neural network models }
\DIFdelend \DIFaddbegin \DIFadd{Machine learning models trained }\DIFaddend with numerical simulation data \DIFdelbegin \DIFdel{mappings}\DIFdelend \DIFaddbegin \DIFadd{can provide a faster alternative to traditional simulators}\DIFaddend . Here we present U-FNO, a novel neural network architecture for solving multiphase flow problems with superior \DIFdelbegin \DIFdel{speed, accuracy, }\DIFdelend \DIFaddbegin \DIFadd{accuracy, speed, }\DIFaddend and data efficiency. U-FNO is designed based on the newly proposed Fourier neural operator (FNO)\DIFdelbegin \DIFdel{that learns an infinite dimensional integral kernel in the Fourier space}\DIFdelend , which has shown excellent performance \DIFdelbegin \DIFdel{for }\DIFdelend \DIFaddbegin \DIFadd{in }\DIFaddend single-phase flows. \DIFdelbegin \DIFdel{Here we }\DIFdelend \DIFaddbegin \DIFadd{ We }\DIFaddend extend the FNO-based architecture to a \DIFdelbegin \DIFdel{CO$_2$-water multiphase problem , and propose the }\DIFdelend \DIFaddbegin \DIFadd{highly complex CO$_2$-water multiphase problem with wide ranges of permeability and porosity heterogeneity, anisotropy, reservoir conditions, injection configurations, flow rates, and multiphase flow properties. The }\DIFaddend U-FNO architecture \DIFdelbegin \DIFdel{ to enhance the prediction accuracy in multiphase flow systems. Through a systematic comparison among a }
\DIFdel{CNN}
\DIFdel{benchmark and three types of FNO variations, we show that the }\DIFdelend \DIFaddbegin \DIFadd{ is more accurate in gas saturation and pressure buildup predictions than the original FNO and a state-of-the-art convolutional neural network (CNN) benchmark. Meanwhile, it has superior data utilization efficiency, requiring only a third of the training data to achieve the equivalent accuracy as CNN. }\DIFaddend \DIFdelbegin \DIFdel{U-FNO architecture has the advantages of both the traditional CNN and original FNO, providing significantly more accurate and efficient performance than previous architectures. The trained U-FNO predicts }\DIFdelend \DIFaddbegin \DIFadd{U-FNO provides superior performance in highly heterogeneous geological formations and critically important applications such as }\DIFaddend gas saturation and pressure buildup \DIFdelbegin \DIFdel{with a $6\times10^4$ times }
\DIFdel{speed-up}
\DIFdel{compared to traditional numerical simulators while maintaining similar accuracy}\DIFdelend \DIFaddbegin \DIFadd{“fronts” determination}\DIFaddend . The trained \DIFdelbegin \DIFdel{models provide }\DIFdelend \DIFaddbegin \DIFadd{model can serve as }\DIFaddend a general-purpose \DIFdelbegin \DIFdel{substitute for }\DIFdelend \DIFaddbegin \DIFadd{alternative to }\DIFaddend routine numerical simulations of 2D-radial CO$_2$ injection problems with \DIFdelbegin \DIFdel{wide ranges of permeability and porosity heterogeneity, anisotropy, reservoir conditions, injection configurations, flow rates, and multiphase flow properties}\DIFdelend \DIFaddbegin \DIFadd{significant speed-ups than traditional simulators}\DIFaddend .
\end{abstract}




\end{frontmatter}


\section{Introduction}
\label{sec:intro}
Multiphase flow in porous media is important for many geoscience applications, including contaminant transport~\citep{bear2010modeling}, carbon capture and storage (CCS)~\citep{pachauri2014climate}, hydrogen storage~\citep{hashemi2021pore}, oil and gas extraction~\citep{aziz1979petroleum}, and nuclear waste storage~\citep{amaziane2012numerical}. Due to the multi-physics, non-linear, and multi-scale nature of these processes, numerical simulation is the primary approach used to solve mass and energy conservation equations for these applications~\citep{orr2007theory}. These numerical simulations are often very time consuming and computationally intensive since they require fine spatial and temporal discretization to accurately capture the flow processes~\citep{Doughty2010, Wen2019}. Meanwhile, the inherent uncertainty in property distributions of heterogeneous porous \replaced{media}{medium} necessitates probabilistic assessments and inverse modeling to aid engineering decisions~\citep{Kitanidis2015, Strandli2014}. Both of these procedures require large numbers of forward numerical simulation runs and are often prohibitively expensive~\citep{NationalAcademiesofSciencesEngineering2018}.

A number of machine learning-based methods have been proposed over the past few years to provide faster alternatives to numerical simulation~\citep{tahmasebi2020machine}. Most existing machine learning-based methods can be categorized into the following two categories: (1) data-driven finite-dimensional operators that learn Euclidean space mappings from numerical simulation data~\citep{zhu2018bayesian, mo2019deep, Zhong2019, tang2020deep, WEN2021103223, WEN2021104009}, and (2) physics-informed/ physics-constrained/neural finite difference learning methods that parameterize the solution functions with a neural network~\citep{raissi2019physics, Zhu2019, haghighat2021sciann}. The first type, finite-dimensional operators, is often implemented with convolutional neural networks (CNN). These CNN-based models have been successful in providing fast and accurate predictions for high-dimensional and complex multiphase flow problems~\citep{WEN2021103223, jiang2021deep, WEN2021104009, tang2021deep, wu2020physics}. However, CNN-based methods are prone to overfitting, therefore requiring large numerical simulation data sets that can be unmanageable as the problem dimension grows. Also, the results produced by these models are tied to the specific spatial and temporal meshes  used in the numerical simulation data set. The second approach\added{, often implemented with artificial neural networks (ANN) (e.g., CNNs~\citep{wang2021physics, kamrava2021simulating}),} \replaced{uses}{using} neural finite difference methods \added{that }require\deleted{s} separate trainings for any new instance of the parameters or coefficients~\citep{raissi2019physics} (\textit{e.g.}, new permeability map or injection rate). Therefore, these methods require as much computational effort as traditional numerical solvers, if not more. \deleted{Furthermore, these methods often struggle with multiphase flow problems with shock fronts, which are common for many classes of geoscience problems~[25, 26].} \added{Furthermore, for the Buckley-Leverett two-phase immiscible flow problem that is common for subsurface flow problems, physics-informed approaches often require observed data or additional diffusive term/physical constraint to improve convergence}~\citep{fuks2020physics, almajid2021prediction, fraces2021physics}.

Recently, a novel approach, the \textit{neural operator}, has been proposed that directly learns the infinite-dimensional-mapping from any functional parametric dependence to the solution~\citep{lu2019deeponet, li2020multipole, li2020neural, bhattacharya2020model}. Unlike neural finite difference methods, neural operators are data-driven therefore require training only once. Meanwhile, neural operators are mesh-independent, so they can be trained and evaluated on different grids. Due to the cost of evaluating global neural integral operators, previously proposed neural operators have not yet been able to achieve the desirable degree of computational efficiency~\citep{li2020fourier}. However, one type of neural operator, the Fourier neural operator (FNO), alleviates this issue through the implementation of a Fast Fourier Transform~\citep{li2020fourier}. The FNO has shown excellent performance on single-phase flow problems with great generalization ability, and is significantly more data efficient than CNN-based methods~\citep{li2020fourier}. 

Here we extend the FNO-based architecture to multiphase flow problems. We find that while FNO's testing accuracy is generally higher than CNN-based models, the training accuracy is sometimes lower due to the regularization effect of the FNO architecture. To improve upon this, we present an enhanced Fourier neural operator, named U-FNO, that combines the advantages of FNO-based and CNN-based models to provide results that are both highly accurate and data efficient. Through the implementation of the newly proposed U-Fourier layer, we show that the U-FNO model architecture produces superior performance over both the original FNO~\citep{li2020fourier} and a state-of-the-art CNN benchmark~\citep{WEN2021104009}. We apply the U-FNO architecture to the highly complex CO$_2$-and-water multiphase flow problem in the context of CO$_2$ geological storage to predict dynamic pressure buildup and gas saturation. The trained U-FNO models provide an alternative to numerical simulation for 2D-radial CO$_2$ injection problems with wide ranges of permeability and porosity heterogeneity, anisotropy, reservoir conditions, injection configurations, flow rates, and multiphase flow properties.

\section{Problem setting}
\label{sec:materials}
\subsection{Governing equation}
We consider a multi-phase flow problem with CO$_2$ and water in the context of geological storage of CO$_2$. The CO$_2$ and water are immiscible but have mutual solubility. The general forms of mass accumulations for component $\eta=CO_2$ or $water$ are written as~\citep{Pruess1999}:
\begin{align}\label{eqn:9}
\frac{\partial \big(\varphi\sum_pS_p\rho_pX^{CO_2}_p\big)}{\partial t} &=-\nabla\cdot \bigg[\mathbf{F}^{CO_2}|_{adv} + \mathbf{F}^{CO_2}|_{dif}\bigg]+q^{CO_2}  \\
\frac{\partial \big(\varphi\sum_pS_p\rho_pX^{water}_p\big)}{\partial t} &=-\nabla\cdot \bigg[\mathbf{F}^{water}|_{adv} + \mathbf{F}^{water}|_{dif}\bigg].\label{eqn:10}
\end{align}
Here $p$ denotes the phase of $w$ (wetting) or $n$ (non-wetting). In the siliciclastic rocks present at most geological storage sites, water is the wetting phase~\citep{pini2012capillary}. However, due to the mutual solubility of water and CO$_2$, there is a small amount of CO$_2$ in the water phase and a small amount of water in the CO$_2$ phase. Here $\varphi$ is the porosity, $S_p$ is the saturation of phase $p$, and $X_p^\eta$ is the mass fraction of component $\eta$ in  phase $p$.

For both components, the advective mass flux $\mathbf{F}^\eta|_{adv}$ is obtained by summing over phases $p$,
\begin{equation}
\mathbf{F}^\eta|_{adv}=\sum_{p}X^\eta\mathbf{F}_{p}=\sum_{p}X^\eta \big( -k\frac{k_{r,p}\rho_p}{\mu_p}(\nabla P_{p} - \rho_p\mathbf{g}) \big)
\end{equation}
where each individual phase flux $\mathbf{F}_{p}$ is governed by the multiphase flow extension of Darcy's law. $k$ denotes the absolute permeability, $k_{r,p}$ is the relative permeability of phase $p$ that non-linearly depends on $S_p$, $\mu_p$ is the viscosity of phase $p$ that depends on $P_p$, and $\mathbf{g}$ is the gravitational acceleration. 

Due to the effect of capillarity, the fluid pressure $P_p$ of each phase is
\begin{align}
P_n &= P_w + P_{c} \\
P_w &= P_w
\end{align}
where the capillary pressure $P_c$ is a non-linear function of $S_p$. Additionally, porosity $\varphi$, density $\rho_p$, and the solubility of $CO_2$ in Equation~\ref{eqn:9} and Equation~\ref{eqn:10} are also non-linear functions that depend on $P_p$. A table of notation is included in \ref{apx:table}.

To simplify the problem setting, our simulation does not explicitly include molecular diffusion and hydrodynamic dispersion. However some unavoidable  \DIFaddbegin \DIFadd{numerical diffusion and }\DIFaddend numerical dispersion resulting from approximating spatial gradients using the two-point upstream algorithm~\citep{eclipse} is intrinsic to the numerical simulations used for the neural network training.

\subsection{Numerical simulation setting}
We use the numerical simulator ECLIPSE (e300) to develop the multiphase flow data set for CO$_2$ geological storage. ECLIPSE is a full physics simulator that uses the finite difference method with upstream weighting for spatial discretization and the adaptive \replaced{im}{IM}plicit method for temporal discretization~\citep{eclipse}. We inject super-critical CO$_2$ at a constant rate into a radially symmetrical system $x(r,z)$ through a vertical injection well with a radius of 0.1 m. The well can be perforated over the entire thickness of the reservoir or limited to a selected depth interval.  We simulate CO$_2$ injection for 30 years at a constant rate ranging from 0.2 to 2 Mt/year. The  thickness of the reservoir ranges from 12\added{.5} to 200 m with no-flow boundaries on the top and bottom. We use a vertical cell dimension of 2.08 m to capture the vertical heterogeneity of the reservoir. The radius of the reservoir is 100,000 m. The outer boundary is closed, but is sufficiently distant from the injection well that it behaves like an infinite acting reservoir. 

Two hundred gradually coarsened grid cells are used in the radial direction. Grid sensitivity studies show that this grid is sufficiently refined to capture the CO$_2$ plume migration and pressure buildup, while remaining computationally tractable~\citep{Wen2019}. Simulated values of the gas saturation ($SG$) and pressure buildup ($dP$) fields at 24 gradually coarsening time snapshots are used for training the neural nets. Refer to \ref{apx:grid} for detailed spatial and temporal discretizations. 

\begin{figure}[t]
    \centering
    \includegraphics[width=\textwidth]{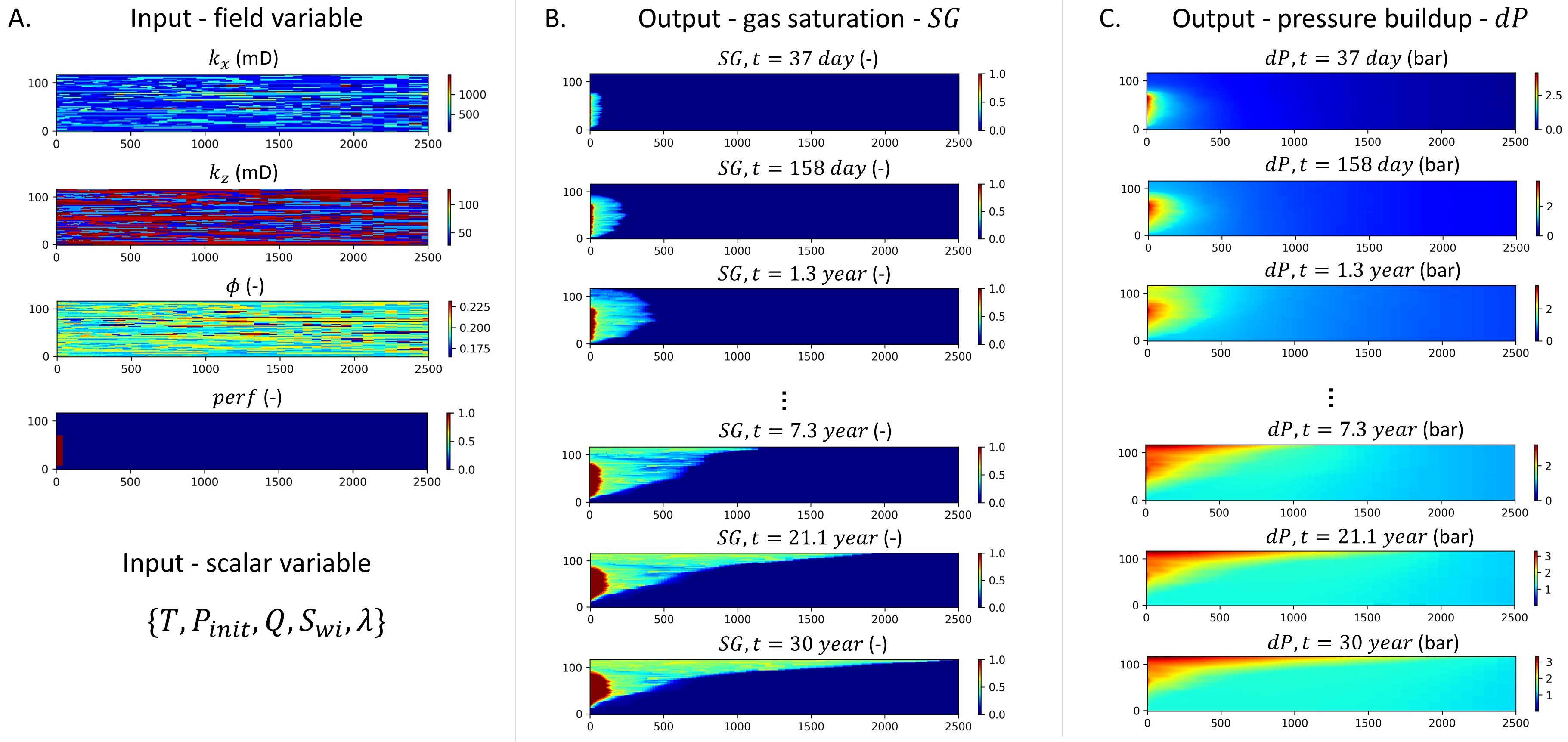}
    \caption{Example of mapping between A. input  to B. output gas saturation and C. pressure buildup. A. Field and scalar channels for each case. Note that the scalar variables are broadcast\deleted{ed} into a channel at the same dimension as the field channels. B. Gas saturation evolution for 6 out of 24 time snapshots. C. Pressure buildup evolution for 6 out of 24 time snapshots.}
    \label{fig:input}
\end{figure}

\subsection{Variable sampling scheme}
We sample two types of variables for each numerical simulation case: field variables and scalar variables. As shown in Figure~\ref{fig:input}, field variables include the horizontal permeability map ($k_x$), vertical permeability map ($k_y$), porosity map ($\phi$), and injection perforation map ($perf$). The reservoir thickness $b$ is randomly sampled in each simulation case and controls the reservoir dimension in each of the following field variables\replaced{. The variable $b$ is applied as an active cell mask to label the rows within the specified thickness.}{:} 

\begin{table}
\footnotesize
\centering
\caption{Summary of input variable's type, sampling range, distribution, and unit. All input sampling are independent with the exception of porosity and vertical permeability map. \added{The dimension of field variables are (96,200).} *: refer to \ref{apx:perm} for a detailed statistical parameter summary for generating heterogeneous $k_x$ map. }
\begin{tabular}{lllll}
 \hline
variable type & sampling parameter & notation & distribution & unit \\ 
\hline
field  & horizontal permeability field & $k_{x}$ & heterogeneous* &-\\ 
& \# of anisotropic materials & $n_{aniso}$ & $X\sim \mathcal{U}\{1,6\}$&-\\
& material anisotropy ratio & $k_{x}/k_{y}$&$X\sim \mathcal{U}[1,150]$&-\\
& porosity (perturbation)                 & $\phi$ &  $\epsilon\sim \mathcal{N}(0,0.005)$&-\\
& reservoir thickness         & $b$     &    $X\sim \mathcal{U}[12$\added{.5}$,200]$&m\\
& perforation thickness & $b_{perf}$ &    $X\sim \mathcal{U}[12,b]$&m\\
& perforation location & - &  randomly placed   & -\\
\hline
scalar & injection rate              & $Q$     &    $X\sim \mathcal{U}[0.2,2]$&MT/y\\
 & initial pressure& $P_{init}$&$X\sim \mathcal{U}[100,300]$&bar\\
& iso-thermal reservior temperature     & $T$ & $X\sim \mathcal{U}[35,170]$&$^{\circ}$C\\
& irreducible water saturation    & $S_{wi}$   &    $X\sim \mathcal{U}[0.1,0.3]$&-\\
& van Genuchten scaling factor    & $\lambda$  &    $X\sim \mathcal{U}[0.3,0.7]$&-\\
\hline
\end{tabular}
\label{tab:1}\end{table}
\begin{itemize}
    \item $k_{x}$: The Stanford Geostatistical Modeling Software (SGeMS)~\citep{sgems} is used to generate the heterogeneous $k_x$ maps. SGeMS produces permeability map according to required input parameters such as correlation lengths in the vertical and radial directions, medium appearances (\ref{apx:perm}), as well as permeability mean and standard deviation. A wide variety of permeability maps representing different depositional environments are included in the data set and the permeability value ranges widely from 10 Darcy to 0.001 mD. \ref{apx:perm} summarizes statistical parameters that characterize the permeability maps. \added{Note that in a radially symmetrical system, these maps form rings of heterogeneity around the injection well. We do not claim these permeability maps are realistic models of any reservoir, but use them to demonstrate the proposed model’s performance in  heterogeneous systems.}
    
    \item $k_{y}$: The vertical permeability map is calculated by multiplying the $k_x$ map by the anisotropy map. To generate the anisotropy map, values of $k_x$ are binned into $n_{aniso}$ materials\replaced{ where each bin}{, each of which} is assigned a randomly sampled anisotropy ratio. \added{The anisotropy ratios are then assigned to the anisotropy map according to the location of the corresponding $k_{x}$.} Note that the anisotropy ratio is uncorrelated with the magnitude of the radial permeability. This procedure roughly mimics a facies-based approach for assigning anisotropy values.
    \item $\phi$: Previous studies show that porosity and permeability are loosely correlated with each other~\citep{pape2000variation}. Therefore, to calculate porosity we first use the fitting relationship presented in Pape et al~\citep{pape2000variation} and then perturb these values with a random Gaussian noise $\epsilon$ with mean value of zero and standard deviation of 0.001.
    \item $perf$: The injection interval thickness $b_{perf}$ is randomly sampled within the range from 12\added{.5} m to the specific reservoir thickness $b$ of that case. We placed the perforation interval on the injection well, by randomly sampling the depth of the perforation top from 0 m to $(b-b_{perf})$ m.
\end{itemize}
Visualizations of the above field variable\added{s} are shown in \ref{apx:perm}. Table~\ref{tab:1} summarizes the parameter sampling ranges and distributions\added{ that are used to generate these field variables}. The sampling parameters are independent of each other with the exception of porosity and permeability. 

Scalar variables include the initial reservoir pressure at the top of the reservoir ($P_{init}$), reservoir temperature ($T$), injection rate ($Q$), capillary pressure scaling factor ($\lambda$)~\citep{li2013influence}, and irreducible water saturation ($S_{wi}$).  The parameter sampling range and distributions are summarized in Table \ref{tab:1}. While the scalar variables $P_{init}$ and $T$ and determined independently, cases that yield unrealistic combinations of these variables are excluded. These field and scalar input variables create a very high-dimensional input space, which\replaced{ often requires massive training data to avoid overfitting when using}{is very challenging for} traditional CNN-based models. \deleted{Nevertheless, the U-FNO model architecture handles the high-dimensional input space with excellent data efficiency.}

\section{Methods}
\label{sec:methods}
\begin{figure}[h]
    \centering
    \includegraphics[width=\textwidth]{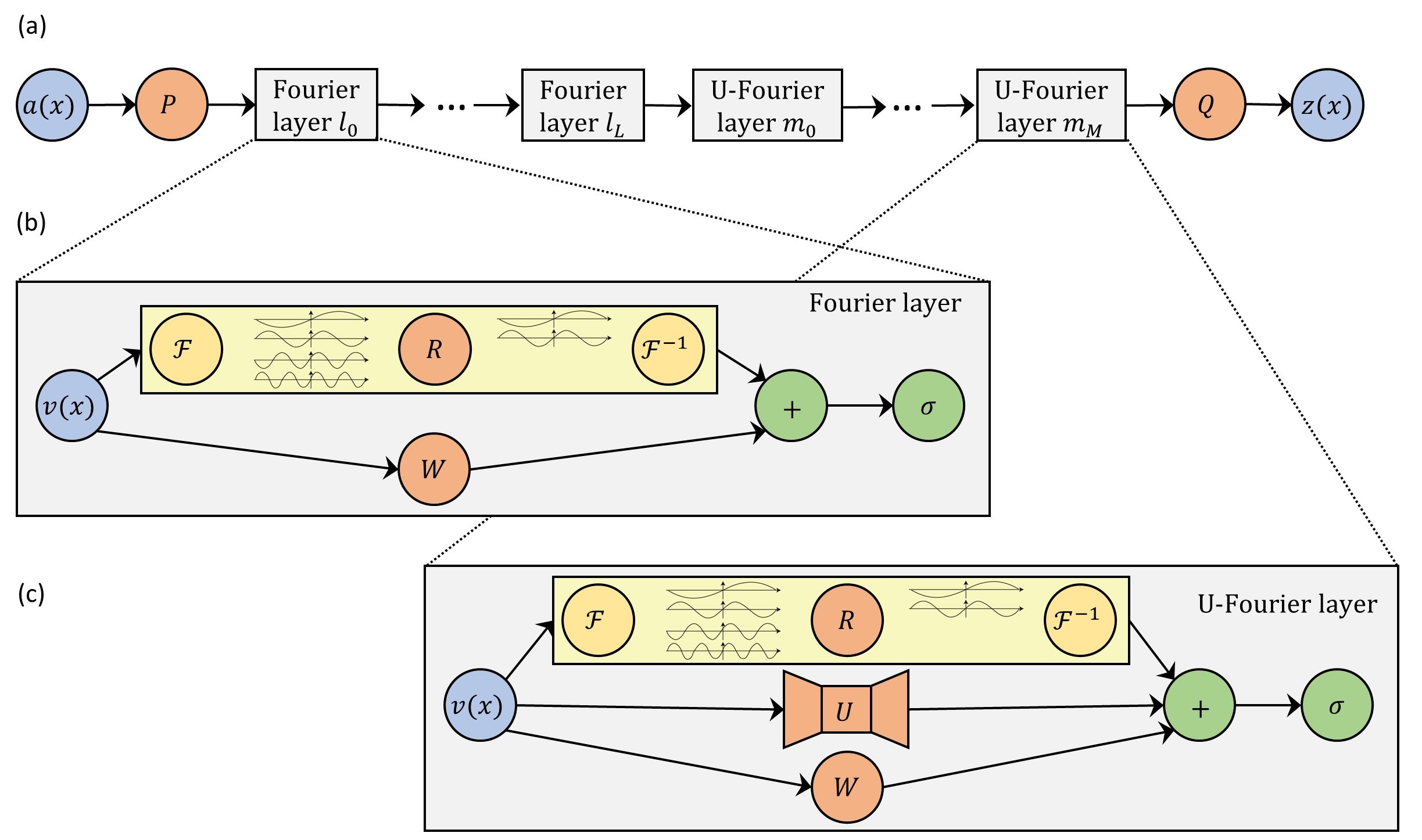}
    \caption{A. U-FNO model architecture. $a(x)$ is the input, $P$ and $Q$ are fully connected neural networks, and $z(x)$ is the output. B. Inside the Fourier layer, $\mathcal{F}$ denotes the Fourier transform, $R$ is the parameterization in Fourier space, $\mathcal{F}^{-1}$ is the inverse Fourier transform, $W$ is a linear bias term, and $\sigma$ is the activation function. C. Inside the U-FNO layer, $U$ denotes a two step U-Net, the other notations have identical meaning as in the Fourier layer.}
\label{fig:model}\end{figure}

The goal of a neural operator is to learn an infinite-dimensional-space mapping from a finite collection of input-output observations. To formulate the problem, we define the domain $D\subset \mathds{R}^d$ be a bounded and open set; $\mathcal{A}$ be the input function space; $\mathcal{Z}$ be the output function space. $\mathcal{A}$ and $\mathcal{Z}$ are separable Banach spaces of functions defined on $D$ that take\deleted{s} values in $\mathds{R}^{d_a}$ and $\mathds{R}^{d_z}$ respectively.   $\mathcal{G}^\dag:\mathcal{A}\to \mathcal{Z}$ is a non-linear map that satisf\replaced{ies}{y} the governing PDEs. Suppose we have $a_j$ that are drawn from probability measure $\mu$ in $\mathcal{A}$, then $z_j=\mathcal{G}^\dag(a_j)$. We aim to build an operator $\mathcal{G}_\theta$ that learns an approximation of $\mathcal{G}^\dag$ by minimizing the following problem using a cost function $C$.
\begin{equation}
    \min_\theta \mathds{E}_{a\sim \mu}[C(\mathcal{G}_{\theta}(a),\mathcal{G}^\dag(a))]
\end{equation}
Since $a_j\in\mathcal{A}$ and $z_j\in\mathcal{Z}$ are both functions, we use $n$-point discretization $D_j=\{x_1,...,x_n\}\subset D$ to numerically represent $a(x)_j|_{D_j}\in \mathds{R}^{n \times d_a}$ and $z(x)_j|_{D_j}\in \mathds{R}^{n \times d_z}$. \added{In this paper, $a(x)_j$ represents the field and scalar variables described in Section 2.3; $z(x)_j$ represents the outputs of temporally varying gas saturation and pressure buildup fields. }We demonstrate in this section that the proposed U-FNO architecture learns the infinite-dimensional-space mapping $\mathcal{G}_\theta$ from a finite collections of $a(x)_j$ and $z(x)_j$ pairs\added{ utilizing integral kernel operators in the Fourier space}. \added{Figure~\ref{fig:model} provides a schematic of the U-FNO architecture. }A table of notation is included in \ref{apx:table}.

\subsection{Integral kernel operator in the Fourier space}
\replaced{We define the}{The} integral kernel operator\added{ (illustrated as the yellow boxes in Figure~\ref{fig:model}b and c)} \deleted{in Equation~\ref{eq:iterative} is defined} by
\begin{equation}\label{eq:2}
    \big( \mathcal{K}(v_l) \big)(x) = \int_D \kappa(x, y)v_l(y)\mathrm{d}v_l(y), \forall_x\in D.
\end{equation}
To efficiently parameterize kernel $\kappa$, the FNO method considers the representation $v_l$ (and also $v_m$) in the Fourier space and utilizes Fast Fourier Transform (FFT)~\citep{li2020fourier}. By letting $\kappa(x,y)=\kappa(x-y)$ in Equation~\ref{eq:2} and applying the convolution theorem, we can obtain 
\begin{equation}
    \big( \mathcal{K}(v_l) \big)(x) = \mathcal{F}^{-1}\big(\mathcal{F}(\kappa)\cdot \mathcal{F}(v_l)\big)(x), \forall_x\in D
\end{equation}
where $\mathcal{F}$ denotes a Fourier transform of a function $f:D\to \mathds{R}^{c}$ and $\mathcal{F}^{-1}$ is its inverse. Now, we can parameterize $\kappa$ directly by its Fourier coefficients: 
\begin{equation}
    \big( \mathcal{K}(v_l) \big)(x) = \mathcal{F}^{-1}\big(R\cdot \mathcal{F}(v_l)\big)(x), \forall_x\in D.
\end{equation}
where $R$ is the Fourier transform of a periodic function $\kappa$. Since we assume that $\kappa$ is periodic, we can apply a Fourier series expansion and work in the discrete modes of Fourier transform. 

We first truncate the Fourier series at a maximum number of modes $k_{max}$, and then parameterize $R$ directly as a complex valued ($k_{max}\times c\times c$)-tensor with the truncated Fourier coefficients. As a result, multiplication by the learnable weight tensor $R$ is
\begin{equation}
    \big(R\cdot \mathcal{F}(v_l)\big)_{k,i}=\sum^{c}_{j=1}R_{k,i,j}(\mathcal{F}(v_l))_{k,j}, \quad\forall k=1,...,k_{max},~ i=1,...,c.
\end{equation}
By replacing the $\mathcal{F}$ by the FFT and implementing $R$ using a direct linear parameterization, we have obtained the Fourier operator as illustrated in Figure~\ref{fig:model}B and C with nearly linear complexity.

\subsection{U-FNO architecture}
The U-FNO architecture contains the following three steps:
\begin{enumerate}
    \item Lift input observation $a(x)$ to a higher dimensional space $v_{l_0}(x)=P(a(x))$ through a fully connected neural network transformation $P$.
    \item Apply iterative Fourier layers followed by iterative U-Fourier layers: $v_{l_0}\mapsto  ... \mapsto v_{l_L} \mapsto v_{m_0}\mapsto ... \mapsto v_{m_M}$ where $v_{l_j}$ for $j=0,1,...,L$ and $v_{m_k}$ for $k=0,1,...,M$\deleted{-1} are sequences of functions taking values in $\mathds{R}^{c}$ for channel dimension $c$.
    \item Project $v_{m_M}$ back to the original space $z(x)=Q(v_{m_M}(x))$ using a fully connected neural network transformation $Q$.
\end{enumerate}

\deleted{Figure~\ref{fig:model}A provides a schematic of the U-FNO architecture. }Within each newly proposed U-Fourier layer~\added{(Figure 2c)}, we have
\begin{equation}
    v_{m_{k+1}}(x):=\sigma\bigg(\big( \mathcal{K}v_{m_{k}} \big)(x)+ \big(\mathcal{U}v_{m_{k}}\big)(x)+W (v_{m_{k}}(x))\bigg) ,\forall x \in D
\label{eq:iterative}\end{equation}
where $\mathcal{K}$ is \replaced{the}{a} kernel integral transformation \replaced{defined above}{parameterized by a neural network}, $\mathcal{U}$ is a U-Net CNN operator, and $W$ is a linear operator, which are all learnable. $\sigma$ is \replaced{an}{a non-linear} activation function\added{ that introduces strong non-linearity to each U-Fourier layer}. Refer to Li et al.~\citep{li2020fourier} for the formulation of the original Fourier layer.

\subsection{Characteristics of the U-Fourier layer}
In contrast to the original Fourier layer in FNO~\citep{li2020fourier}, the U-FNO architecture proposed here appends a U-Net path in each U-Fourier layer. The U-Net processes local convolution to enrich the representation power of the U-FNO in higher frequencies information. The number of Fourier and U-Fourier layers, $L$ and $M$, are hyperparameters that can be optimized for the specific problem. For the multi-phase flow problem considered here, we found that the architecture with half Fourier layers and half U-Fourier layers achieves the best performance, compared to architectures with all Fourier layers or all U-Fourier layers. 

Note that\deleted{ although} the Fourier neural operator is an infinite-dimensional-operator, \added{which generates mesh-free/resolution invariant predictions. However, } when we append the U-Net block, we \replaced{introduced the CNN-based path that does not inherently provide}{sacrifice} the flexibility of training and testing\deleted{ the model} at different discretizations. We made this choice because the CO$_2$-water multiphase flow problem is very sensitive to numerical dispersion and numerical \replaced{dissolution}{diffusion}, which are both tied to a specific grid resolution. When training and testing at different grid dimensions, the numerical noise is often transformed in a nonphysical way. As a result, for this problem, we prioritize achieving higher training and testing accuracy, which the U-FNO provides. \added{Nevertheless, under the circumstance where one wants to test the U-FNO model at unseen grid resolutions, we developed additional down-sampling and up-sampling techniques that can be applied to the U-Net component to re-introduce the resolution invariant feature. An example of this technique and its performance are discussed in Section 5.3.}

\added{Finally, the U-Fourier layer's performance improvement is not limited to the spatial-temporal 3D multiphase flow problem considered in this paper. We found that the U-FNO's 2D variation also outperforms the original FNO-2D in a steady-state Darcy's flow problem. Refer to \ref{apx:darcy} for details.}

\subsection{Data configuration}
\added{This section describes the configuration of the inputs and outputs for the proposed U-FNO architecture. For the data input, e}\deleted{E}ach of the field variables in Figure~\ref{fig:input}A is represented by a channel\deleted{ in the data input}. Since we use a gradually coarsening radial grid for the numerical simulations, a logarithm conversion in the radial direction is applied in training to project the field variables onto a uniform grid that can be represented by a $(96,200)$ matrix. Notice that reservoir thickness is also a variable and 96 cells represents a 200 m thick reservoir. When the reservoir is thinner than 200 m, we use zero-padding to denote cells that are outside of the actual reservoir. For the scalar variables, the values are simply broadcast into a matrix with dimension of $(96,200)$.

In addition to the input variables, we also supply the spatial grid information to the training by using one channel to denote radial cell dimensions and another channel to denote vertical cell dimensions. The temporal grid information is supplied into the network as an additional dimension. The input to each data sample is constructed by concatenating the field variables, scalar variables, spatial grids, and temporal grid together.

For the gas saturation and pressure buildup outputs as shown in Figure~\ref{fig:input}B and C, we use the same logarithm conversion to project the outputs onto a uniform grid. We then concatenate the outputs for different time snapshots to obtain a spatial-temporal 3D volume. The pressure buildup is normalized into zero-mean and unit-variance distribution. For gas saturation, we do not normalize the data because the saturation values always range from 0 to 1. The dimensions of the input and outputs are shown for in each model architecture (Appendices D to G).

The data set contains 5,\replaced{5}{0}00 input-to-output mappings.  We use a 9/1\added{/1} split to segregate the data set into  4,500 samples for training\added{, 500 samples for validation,} and  500 samples for testing. 

\subsection{Loss function design and training}
We use a relative $lp$-loss to train the deep learning models. The $lp$-loss is applied to both the original output ($y$ \added{$(r,z,t)$}) and the first derivative of the output in the $r$-direction ($\sfrac{\mathrm{d}y}{\mathrm{d}r}$), and is written as:
\begin{equation}
    L(y,\hat{y}) = \frac{||y-\hat{y}||_p}{{||y||_p}}+\beta \frac{||\sfrac{\mathrm{d}y}{\mathrm{d}r}-\hat{\sfrac{\mathrm{d}y}{\mathrm{d}r}}||_p}{{||\sfrac{\mathrm{d}y}{\mathrm{d}r}||_p}},
\label{eq:loss}\end{equation}
where $\hat{y}$ is the predicted output, $\hat{\sfrac{\mathrm{d}y}{\mathrm{d}r}}$ is the first derivative of the predicted output, $p$ is the order of norm, and $\beta$ is a hyper-parameter. This relative loss has a regularization effect and is particularly effective when the data have large variances on the norms. Our experiments show that, compared to an $MSE$-loss, a relative loss significantly improves the performance for both gas saturation and pressure buildup. The second term in Equation~\ref{eq:loss} greatly improves quality of predictions for gas saturation at the leading edge of the plume. Similarly  this term improves prediction of the sharp pressure buildup around the injection well. We use  the $l2$-loss for gas saturation and pressure buildup since it provides faster convergence than the $l1$-loss.

As described in Section \replaced{2}{3}, our data set contains reservoirs with various thicknesses and the cells outside of the reservoir are padded with zeros for both input and output. To accommodate for the variable reservoir thicknesses, during training, we construct an active cell mask for each data sample and only calculate the loss within the mask. Our experiments show that this loss calculation scheme achieves better performance than calculating the whole field because of the better gradient distribution efficiency. 

During training, the initial learning rate is \added{specified to be }0.001 and the learning rate gradually decreases with a constant step and reduction rate. These hyper-parameters are optimized for the gas saturation and pressure buildup model separately. The training stops when the loss no longer decreases\added{, which is 100 and 140 epochs for the gas saturation and pressure buildup model respectively}.

\section{Results}
This section compares 4 types of model architectures: original FNO proposed in Li et al.~\citep{li2020fourier}, the newly proposed U-FNO in this paper, a conv-FNO that uses a \texttt{conv3d} in the place of the U-Net, and the state-of-the-art benchmark CNN used in Wen et al.~\citep{WEN2021104009}. All models are trained on the proposed loss function (Equation~\ref{eq:loss}) and directly output the 3D $(96 \times 200 \times 24)$ gas saturation and pressure field in space and time. Detailed parameters for each model are summarized in Appendices D to G.

\subsection{Gas saturation}

\begin{figure}[h]
    \centering
    \includegraphics[width=\textwidth]{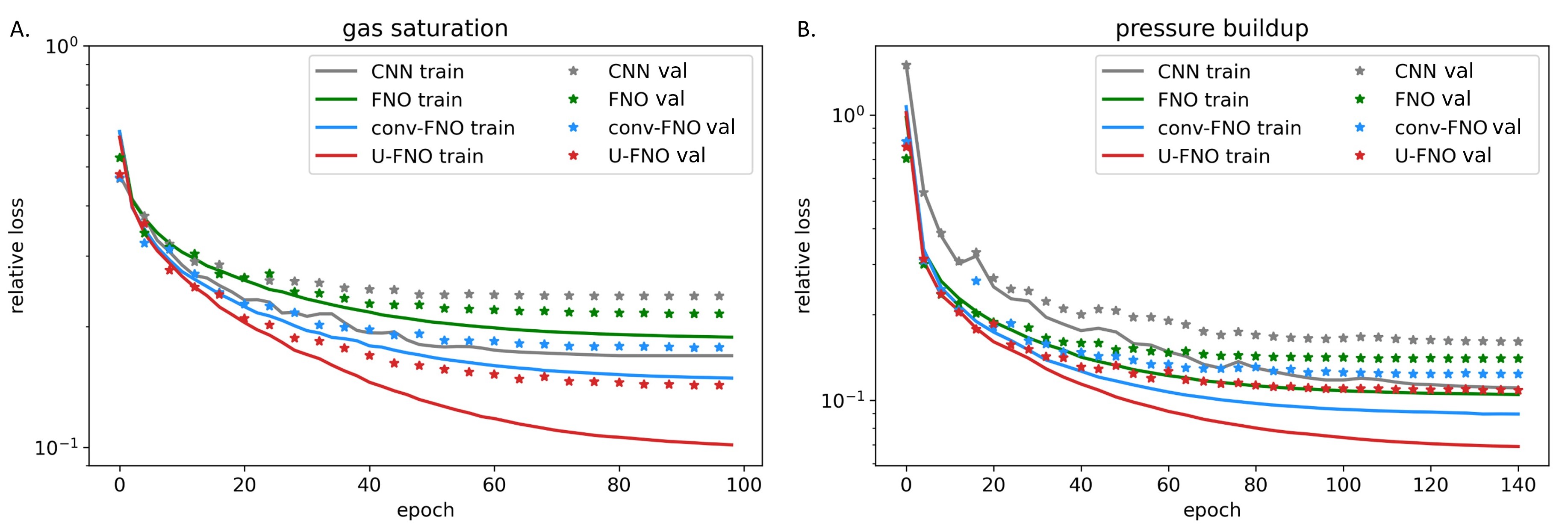}
    \caption{Training and \replaced{validation}{testing} relative loss evolution vs. epoch for U-FNO, FNO, conv-FNO and CNN benchmark for A. gas saturation and B. pressure buildup.}
    \label{fig:loss}
\end{figure}

Figure~\ref{fig:loss}A \deleted{and Table~\ref{tab:r2} }demonstrates that the best performance for both the training and \replaced{validation}{testing} data set is achieved with the U-FNO model. \deleted{Specifically, the testing set performance represents the true predictability of the model for unseen data. The low relative loss clearly indicates the superior performance of the proposed U-FNO. }Interestingly, \added{for the gas saturation model, }we notice that although the original FNO has a higher training relative loss than the CNN benchmark, the \replaced{validation}{testing} relative loss by the original FNO is lower than that of the CNN benchmark. This indicates that FNO has excellent generalization ability and achieves better performance than the CNN even though FNO has a higher training relative loss. Nevertheless, the original FNO has the highest relative loss in the training set due to the inherent regularization effect by using a finite set of truncated Fourier basis. The Conv-FNO and U-FNO architecture is therefore designed to enhance the expressiveness by processing the higher frequency information that are not picked up by the Fourier basis. We can observe from Figure~\ref{fig:loss}A that the training loss is significantly improved even by simply adding a plain \texttt{conv3d} in the Conv-FNO case. When the FNO layer is combined with a U-Net in the U-FNO case, the model takes the advantages of both architectures and consistently produces the lowest relative loss throughout the entire training (Figure~\ref{fig:loss}A).

\begin{figure}[h]
    \centering
    \includegraphics[width=\textwidth]{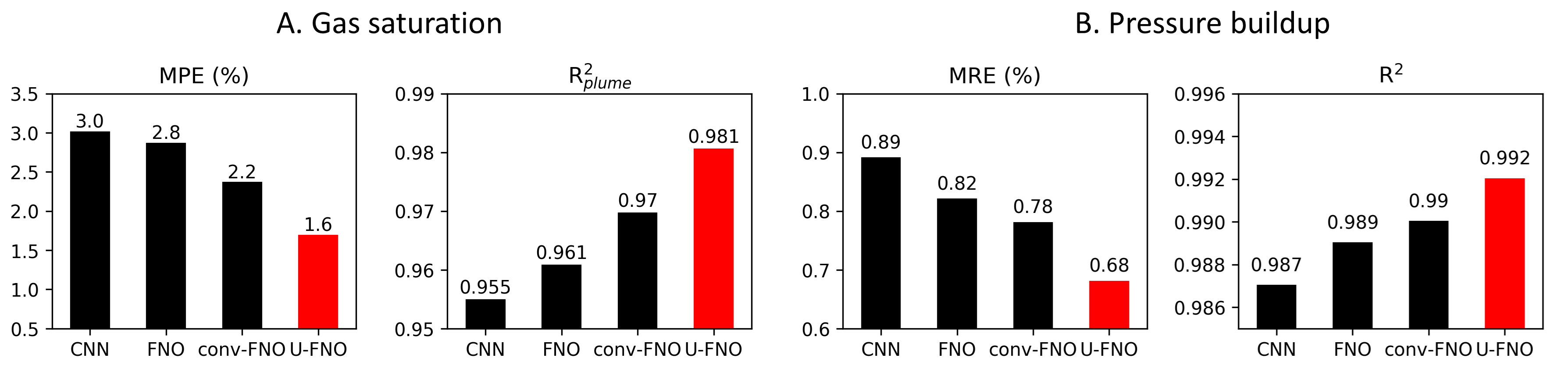}
    \caption{A. Gas saturation testing set plume mean absolute error ($MPE$) and plume $R^2$ scores ($R^2_{plume}$) using CNN, FNO, conv-FNO, and U-FNO. B. Pressure buildup field mean relative error ($MRE$) and $R^2$ scores using the same four models.}
    \label{fig:accuracy}
\end{figure}

\DIFdelbegin \DIFdel{In Table 2(a)}\DIFdelend \DIFaddbegin \DIFadd{Figure~\ref{fig:accuracy}A demonstrates the }\DIFaddend \DIFdelbegin \DIFdel{we summarized the training, validation, and testing set performances evaluated with the average field mean absolute error, }\DIFdelend \DIFaddbegin \DIFadd{testing set }\DIFaddend plume mean absolute error \DIFdelbegin \DIFdel{, and }\DIFdelend \DIFaddbegin \DIFadd{($MPE$) and plume }\DIFaddend $R^2$ scores \DIFaddbegin \DIFadd{($R^2_{plume}$) for each model architectures. }\DIFaddend \DIFaddbegin \DIFadd{We evaluate the gas saturation models' accuracy within the CO$_2$ separate phase plume because the gas saturation outside of the plume is always 0. Here ``within the plume" is defined as non-zero values in either data or prediction. }\DIFaddend The testing set results represent the predictability of the model on truly unseen data \DIFdelbegin \DIFdel{, which further demonstrates the superior performance of the proposed U-FNO. t }\DIFdelend \DIFaddbegin \DIFadd{ and U-FNO achieves the best performance with the lowest $MPE$ and highest $R^2_{plume}$. Comparing to the benchmark CNN, t}\DIFaddend he average \DIFdelbegin \DIFdel{mean plume error and average plume mean absolute error }\DIFdelend \DIFaddbegin \DIFadd{test set $MPE$ }\DIFaddend using U-FNO is \DIFdelbegin \DIFdel{50\% and }\DIFdelend 46\% lower \DIFdelbegin \DIFdel{respectively}\DIFdelend \DIFaddbegin \DIFadd{ while the $R^2_{plume}$ increased from 0.955 to 0.981. }\DIFaddend We can also compare the degree of overfitting by calculating the difference between the training and testing set \DIFaddbegin \DIFadd{$MPE$ (refer to \ref{Apx:scores} for training set $MPE$)}\DIFaddend. For example, the average \DIFdelbegin \DIFdel{field mean absolute error }\DIFdelend \DIFaddbegin \DIFadd{$MPE$ }\DIFaddend difference in CNN is \DIFdelbegin \DIFdel{53}\DIFdelend \DIFaddbegin \DIFadd{70}\DIFaddend \% higher than in U-FNO (\DIFdelbegin \DIFdel{0.017 and 0.008 }\DIFdelend \DIFaddbegin \DIFadd{1.0\% and 0.3\% }\DIFaddend respectively).

\begin{figure}[h]
    \centering
    \includegraphics[width=\textwidth]{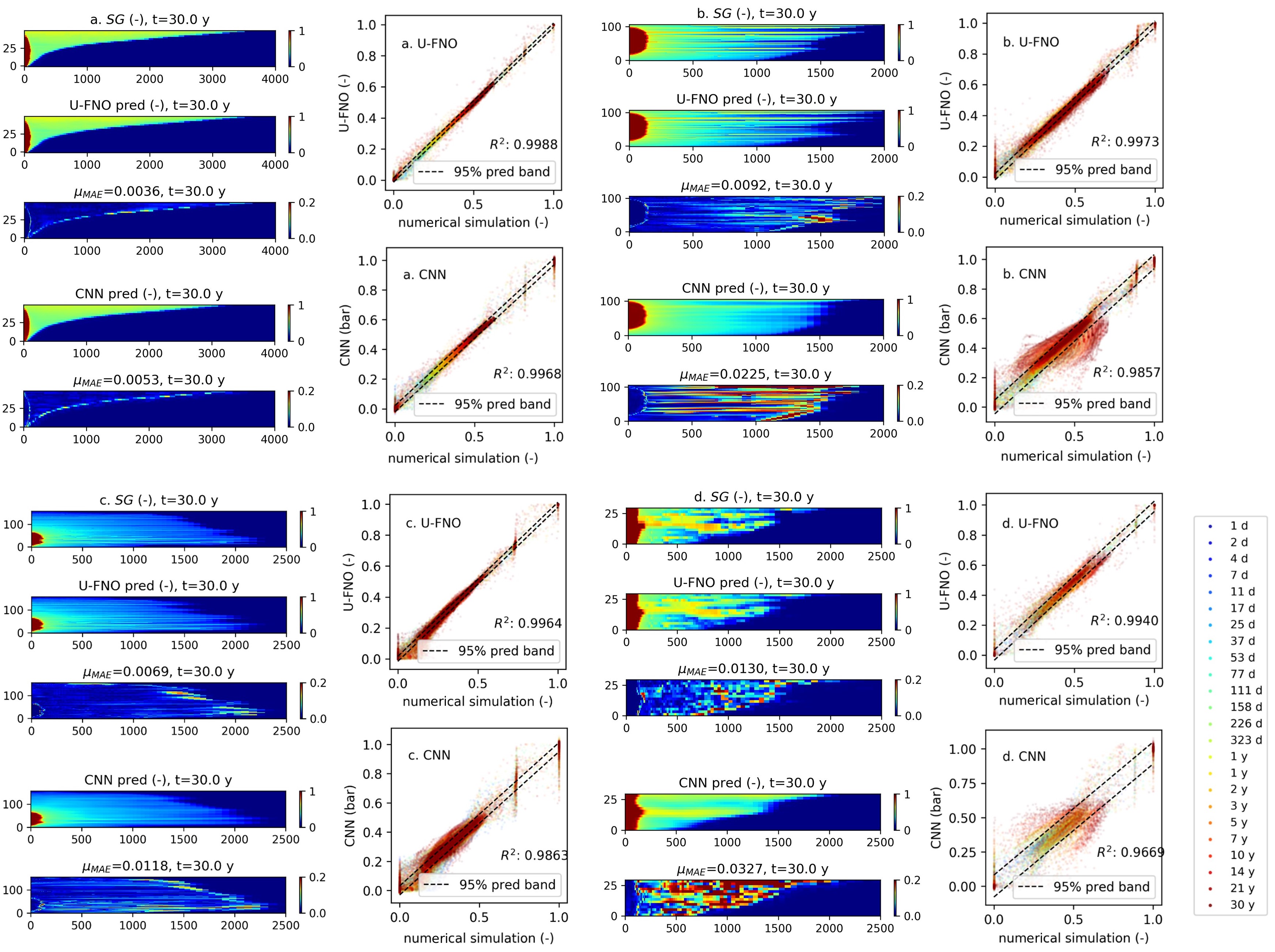}
    \caption{Visualizations and scatter plots for example a to d. In each example, visualizations show the true gas saturation ($SG$), U-FNO predicted, U-FNO absolute error, CNN predicted, and CNN absolute error. The mean absolute error $\mu_{MAE}$ is labeled on the U-FNO and CNN absolute error plots. Scatter plots shows numerical simulation vs. predicted by U-FNO and CNN model on each grid. The legend for all of the scatter plots is shown in the bottom right.}
    \label{fig:sg}
\end{figure}

In addition to considering the average performance over the entire training\added{, validation, }and testing sets, we \DIFaddbegin \DIFadd{also }\DIFaddend compare model predictions for four different cases with varying degrees of complexity in Figure~\ref{fig:sg}. For each case, Figure~\ref{fig:sg} shows a comparison between the predicted and true values of the CO$_2$ saturation for each grid cell in the model over the entire 30 year injection period.  The U-FNO has superior performance compared to the CNN for all of these examples as quantified by the higher $R^2$ value and narrower 95\% prediction bands. Case b. and d. are especially obvious examples in which the U-FNO successfully predicts the complicated horizontal saturation variations where the CNN ignores the heterogeneity and simply predicts more uniform saturation fields.

\subsection{Pressure buildup}

For pressure buildup, the U-FNO also achieves the lowest relative error for both training and \replaced{validation}{testing} data sets. As shown in Figure~\ref{fig:loss}B, the training and \replaced{validation}{testing} relative errors for the U-FNO are consistently low throughout the training process. \DIFdelbegin \DIFdel{Table 2(b) }\DIFdelend \DIFaddbegin \DIFadd{Figure~\ref{fig:accuracy} }\DIFaddend shows U-FNO's superior \DIFaddbegin \DIFadd{testing set }\DIFaddend performance in field mean relative error \DIFaddbegin \DIFadd{($MRE$) }\DIFaddend and $R^2$ score. Specifically, the test set average \DIFdelbegin \DIFdel{field mean absolute error }\DIFdelend \DIFaddbegin \DIFadd{$MRE$ }\DIFaddend is reduced by 24\% from CNN to U-FNO. By comparing the \replaced{differences between}{$R^2$ scores for} the training and testing sets in \DIFdelbegin \DIFdel{Table~\ref{tab:r2}}
\DIFdelend \DIFaddbegin \DIFadd{\ref{Apx:scores}}\DIFaddend , we can observe that all FNO-based models produce \replaced{smaller}{nearly negligible} overfitting\added{ compared to CNN}. 

\begin{figure}[h]
    \centering
    \includegraphics[width=\textwidth]{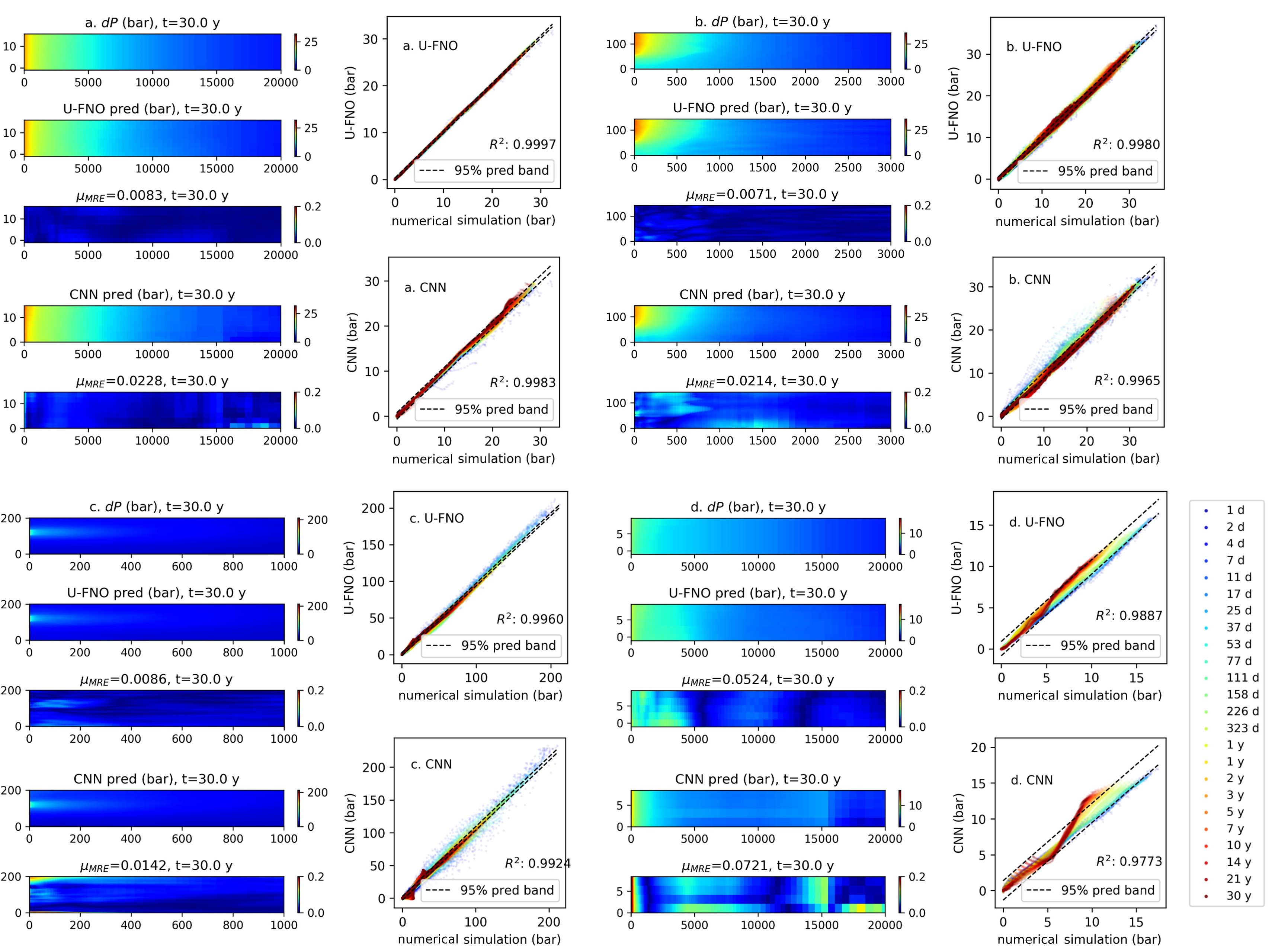}
    \caption{Visualizations and scatter plots for examples a to d. In each example, visualizations show the true pressure buildup ($dP$), U-FNO predicted, U-FNO relative error, CNN predicted, and CNN relative errors. The relative errors are defined as in~\citep{tang2021deep}; the mean relative error $\mu_{MRE}$ is labeled on the U-FNO and CNN relative error plots. Scatter plots shows numerical simulation vs. predicted by U-FNO and CNN model on each grid. The legend for all of the scatter plots is shown in the bottom right.}
    \label{fig:dp}
\end{figure}

The superior performance of the U-FNO for pressure buildup predictions is also demonstrated for the four examples shown in  Figure~\ref{fig:dp}. In each case the U-FNO has higher R$^2$ values and narrower 95\% prediction bands. Unlike the gas saturation outputs, pressure buildup distributions are challenging to predict since they have a larger radius of influence and larger differences between cases. For example, the maximum pressure buildup in the 4 examples shown in Figure~\ref{fig:dp} varies from $\sim$20 bar to $\sim$220 bar. Notice that the the CNN model especially struggles with cases that have large radius of influence (e.g. case d) while the U-FNO model maintains excellent accuracy at locations that are far away from the injection well.

\section{Discussion}
\subsection{U-FNO's advantages over CNN}

\subsubsection{Data utilization efficiency }

\begin{figure}[h]
    \centering
    \includegraphics[width=\textwidth]{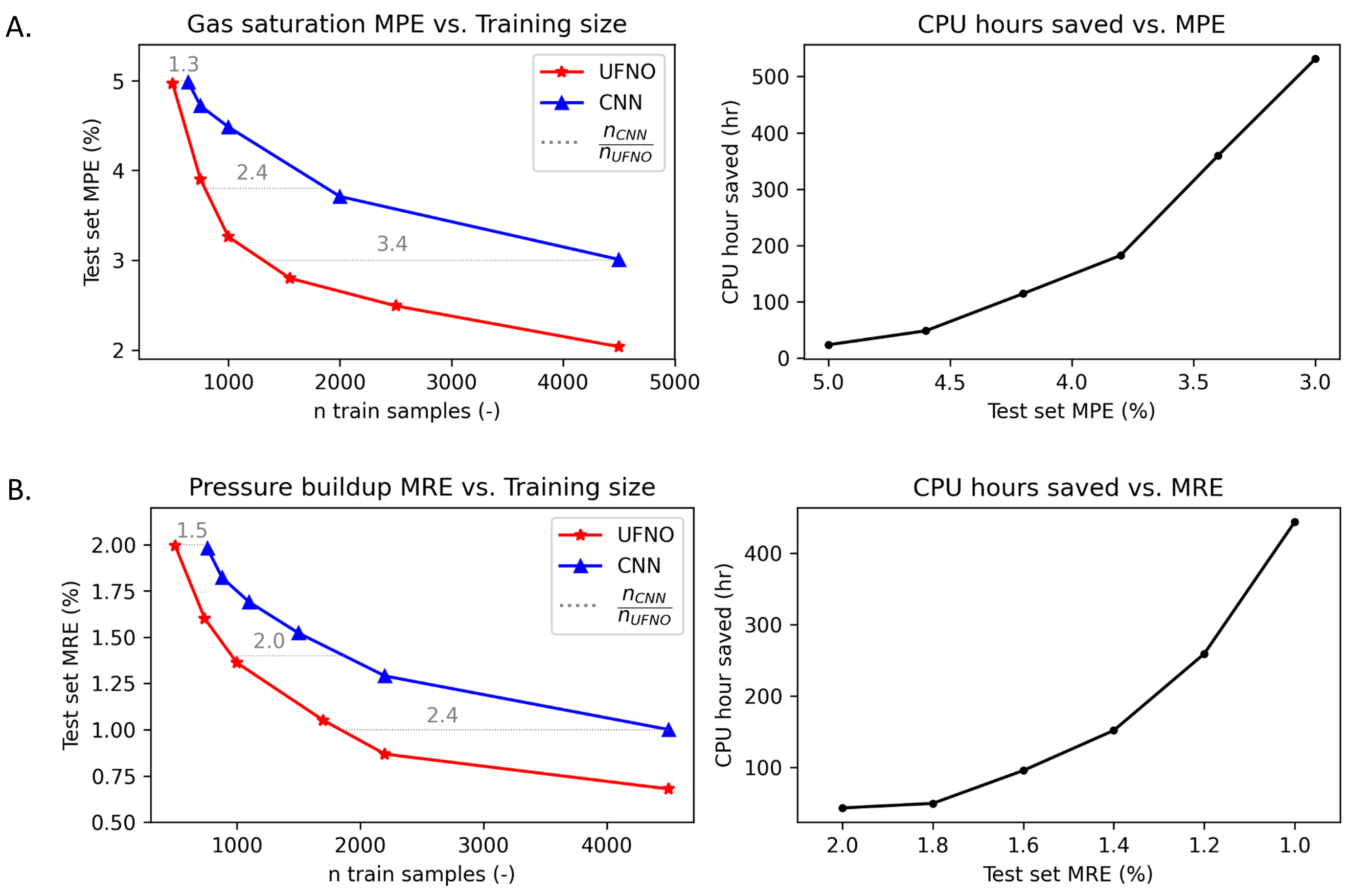}
    \caption{\DIFdelbegin \DIFdel{The training and }
\DIFdel{set $R^2$ mean and standard deviation vs. the number of training sample for A. Gas saturation and B. pressure buildup }\DIFdelend \DIFaddbegin \DIFadd{A. Gas saturation testing set $MPE$ vs. training size for U-FNO and CNN. The grey text labels the CNN to U-FNO training size ratios. The CPU hours saved are calculated using the average simulation time and linear interpolation of the $MPE$ vs. training size relationship. B. Pressure buildup testing set $MRE$ vs. training size for U-FNO and CNN. The CPU hours saved are calculated same as above}\DIFaddend .}
    \label{fig:trainnr}
\end{figure}

The results in Section 4 demonstrate the excellent generalization ability of \DIFdelbegin \DIFdel{all }\DIFdelend \DIFaddbegin \DIFadd{the }\DIFaddend FNO-based architectures. 
To further \DIFdelbegin \DIFdel{investigate the relationship between the training size and overfitting for the }\DIFdelend \DIFaddbegin \DIFadd{compare the data utilization efficiency of the }\DIFaddend newly proposed U-FNO model \DIFdelbegin \DIFdel{, we run a set of comparisons for the U-FNO model trained with 500, 2,500 and 4,500 samples (see }\DIFdelend \DIFaddbegin \DIFadd{with the benchmark CNN, we train each model using various numbers of samples and plotted the testing set $MPE$ and $MRE$ in }\DIFaddend Figure~\ref{fig:trainnr}\DIFdelbegin \DIFdel{)}\DIFdelend . Each model is trained for the same number of epochs.
\DIFdelbegin \DIFdel{From this comparison , we can observe that the gas saturationmodels are more prone to overfitting as indicated by the wider gaps }
\DIFdel{$R^2$ mean and standard deviations. For }\DIFdelend \DIFaddbegin \DIFadd{For gas saturation, the CNN requires up to 3.4 times more training data to achieve the same level of performance as the U-FNO. Similarly, }\DIFaddend the pressure buildup \DIFdelbegin \DIFdel{, the training size of 4,500 produces a very similar $R^2$ score in the training and }
\DIFdel{set. Therefore, a 4,500 }\DIFdelend \DIFaddbegin \DIFadd{CNN requires 2.4 times more training data }\DIFaddend  \DIFdelbegin \DIFdel{data-size is already sufficient for pressure buildup prediction and more training data will not}
\DIFdel{significantly improve the performance}\DIFdelend \DIFaddbegin \DIFadd{ to achieve a test set $MRE$ of 1\%}\DIFaddend .
\DIFaddbegin \DIFadd{In practical terms, the U-FNO saved 530 and 440 CPU hours in data set generation for the gas saturation and pressure buildup models respectively (for a reference CNN model trained with 4500 training samples). Figure~\ref{fig:trainnr} also indicates that the CPU hours saved by using U-FNO grows increasingly as test set errors reduces.
The U-FNO's data utilization efficiency greatly alleviates the computational resource needed in data generation and training, therefore can better support complex high-dimensional problems}\DIFaddend .

\subsubsection{Accuracy in the “front” determination}
\DIFaddbegin \DIFadd{Gas saturation and pressure buildup “fronts” are important quantities for CO$_2$ storage projects and are often used for regulatory oversight~\citep{epa}, monitoring, or history matching~\citep{lengler2010impact} purposes. The distance to the gas saturation “front” corresponds to the maximum extent of the plume of separate phase CO$_2$. The pressure buildup “front” often refers to the radius at a specified threshold value of pressure buildup because pressure fields are smooth. In this experiment, we compare the accuracy of the U-FNO and CNN models to evaluate the gas saturation and pressure buildup “fronts”. Table~\ref{tab:good}(a) and (b) shows that the U-FNO is 2.7 times more accurate than the CNN for saturation “front” prediction and 1.8 times more accurate for pressure “front” prediction. }\DIFaddend 

\begin{table}[h]
\small
\caption{Accuracy of U-FNO and CNN for (a) gas saturation “front” prediction, and (b) pressure buildup “front” prediction. Both comparisons are performed on the testing set.}
    \begin{subtable}[h]{\textwidth}
        \footnotesize
        \centering
        \caption{Here gas saturation “front” is defined as the maximum extend of separate phase CO$_2$ above the threshold value 0.01. The error of gas saturation “front” is calculated as the absolute difference between true and predicted gas saturation “front” divided by true gas saturation “front”.}
        \begin{tabular}{c|cc}
        \hline
                                    & CNN     & U-FNO  \\\hline
    Gas saturation “front” error (\%)  & 9.2    & 3.4   \\
    \hline
       \end{tabular}
    \end{subtable}

    \vspace*{0.3 cm}
    
    \begin{subtable}[h]{\textwidth}
    \footnotesize
        \centering
        \caption{Here pressure buildup “front” is defined as the radius of pressure buildup above the threshold value 0.5 bar. The error of pressure buildup “front” is calculated same as in (a).}
        \begin{tabular}{c|cc}
        \hline
                                  & CNN     & U-FNO  \\\hline
            Pressure buildup “front” error (\%) & 21.2   & 12.0  \\
            \hline
        \end{tabular}
        \end{subtable}
     
     \vspace*{0.3 cm}
\label{tab:good}
\end{table}

\subsubsection{Accuracy in the heterogeneous geological formations}
\DIFaddbegin \DIFadd{The U-FNO is more accurate than the CNN for highly heterogeneous geological formations. The training data set includes a wide variety of homogeneous to heterogeneous permeability maps. For this comparison, we selected the most “heterogeneous” and most “homogeneous” formations from the testing set that have the highest and lowest 10\% permeability standard deviations. Table~\ref{tab:heteo} summarized the average gas saturation $MPE$ in both types of formations using U-FNO and CNN. For the most heterogeneous geological formations, the U-FNO is 1.7 times more accurate than CNN in gas saturation.}\DIFaddend

\begin{table}[h]
\centering
\footnotesize
    \caption{Gas saturation $MPE$ in the most heterogeneous and homogeneous formations with the highest and lowest 10\% permeability  standard deviations in the testing set.}
    \begin{tabular}{c|cc}
         \hline
            Gas saturation $MPE$ (\%)   & CNN     & U-FNO      \\\hline
            “Heterogeneous” formation & 4.7    & 2.7   \\
            “Homogeneous” formation   & 2.0    & 1.5  \\\hline
         \end{tabular}\label{tab:heteo}
 \end{table}

\subsection{Computational efficiency analysis}
We summarize the computational efficiency of the CNN, FNO, Conv-FNO, and U-FNO in Table~\ref{tab:efficiency}. The training and testing times are both evaluated on a Nvidia A100-SXM GPU. Once the gas saturation and pressure buildup models are trained, we can directly \DIFdelbegin \DIFdel{infer }\DIFdelend \DIFaddbegin \DIFadd{use }\DIFaddend these deep learning models \DIFdelbegin \DIFdel{to obtain needed multi-phase flow outputs instead of running the forward numerical simulator}\DIFdelend \DIFaddbegin \DIFadd{as a general-purpose numerical simulator alternative~\citep{WEN2021104009}}\DIFaddend . \DIFdelbegin \DIFdel{as long as the inputs are within the training range of the models.
In our previous work, we demonstrated the viability of using trained deep learning models and a web application as a numerical simulator alternative~\mbox{
\citep{WEN2021104009}}\hspace{0pt}
. The models proposed in this paper can serve in similar ways to support downstream use cases}\DIFdelend \DIFaddbegin \DIFadd{Note that when machine learning models are used in a task-specific “surrogate" context, the training and data collection time are sometimes included in the computational efficiency calculation. However, for the application that we are proposing, the model is trained only once. For subsequent predictions, the trained machine learning model is directly used. Therefore, we compare the machine learning model prediction time to the time that would have been required by using the numerical simulator. }\DIFaddend \DIFdelbegin \DIFdel{Therefore, to }\DIFdelend \DIFaddbegin \DIFadd{To }\DIFaddend evaluate the\added{ computational efficiency speed-up}\deleted{ comparison}, we compare the forward \DIFdelbegin \DIFdel{numerical }\DIFdelend simulation CPU run time \DIFdelbegin \DIFdel{with the deep learning model}\DIFdelend \DIFaddbegin \DIFadd{with each machine learning model's }\DIFaddend testing time. We run ECLIPSE simulations on an Intel\textsuperscript{®} Xeon\textsuperscript{®} Processor E5-2670 CPU. Each simulation \DIFdelbegin \DIFdel{case can utilize }\DIFdelend \DIFaddbegin \DIFadd{uses }\DIFaddend a fully dedicated CPU\DIFdelbegin \DIFdel{and the }\DIFdelend \DIFaddbegin \DIFadd{. The }\DIFaddend average run time for 1,000 random cases is 10 minutes\added{ per run. \DIFdelbegin \DIFdel{We run the comparison with the CPU that's readily available to us. Other types of CPU will note change the simulation time for a significant order}\DIFdelend \DIFaddbegin \DIFadd{Faster CPUs are available but will not materially change the result of this analysis}\DIFaddend .}

\begin{table}[h]
\centering
\scriptsize
\caption{Summary of the number of parameters, training time, and testing times required for all four models. The testing times are calculated by taking the average of 500 random cases. \added{The gas saturation and pressure models can be tested separately. }The speed-up is compared with average numerical simulation run time of 10 mins.}
\begin{tabular}{c|c|c|ccc}
\hline
       & \multirow{2}{*}{\# Parameter} &\multirow{2}{*}{Training} & \multicolumn{3}{c}{Testing}\\
       \cline{4-6}
       & & & Gas saturation & Pressure  & Speed-up vs. numerical  \\
       & (-) & (s/epoch) &  (s) & buildup (s) & simulation (times) \\
\hline
CNN      & 33,316,481 & 562 & 0.050  & 0.050  & 1\replaced{$\times10^4$}{$\times10^5$}         \\
FNO      & 31,117,541 & 711 &   0.005  &  0.005   &   1\replaced{$\times10^5$}{$\times10^6$}           \\
Conv-FNO & 31,222,625 & 1,135 &   0.006  &  0.006  &   1\replaced{$\times10^5$}{$\times10^6$}           \\
U-FNO    & 33,097,829 & 1,872 &   0.010  &  0.010   &   6\replaced{$\times10^4$}{$\times10^5$}         \\
\hline
\end{tabular}
\label{tab:efficiency}
\end{table}

\DIFdelbegin \DIFdel{All }\DIFdelend \DIFaddbegin \DIFadd{Gas saturation and pressure buildup predictions made with all }\DIFaddend of the neural network models are at least 10$^4$ times faster than conventional numerical simulation\DIFdelbegin \DIFdel{during prediction}\DIFdelend . Notice that FNO-based models are significantly faster \DIFdelbegin \DIFdel{than the CNN model at testing time while }\DIFdelend \DIFaddbegin \DIFadd{at testing but }\DIFaddend slower at training \DIFdelbegin \DIFdel{time}\DIFdelend \DIFaddbegin \DIFadd{than the CNN model}\DIFaddend . \DIFdelbegin \DIFdel{In this problem }\DIFdelend \DIFaddbegin \DIFadd{For our problem}\DIFaddend , we prioritize the prediction accuracy and testing time over the training time, which \DIFaddbegin \DIFadd{the }\DIFaddend U-FNO provides. For problems that are more sensitive to training time, one could also use the Conv-FNO which provides both high accuracy and relatively fast training. 

\subsection{Inference at unseen time steps}
\added{FNO-based architectures are infinite-dimensional operators that can provide grid-invariant predictions. However, by adding the convolution path in conv-FNO and the U-Net path in U-FNO, we sacrificed the inherent grid-invariant feature of the FNOs. To reintroduce the ability to provide predictions at unseen time steps, for conv-FNO and U-FNO, we applied additional down-sampling and up-sampling operations to the convolution blocks and U-Net blocks, which transform the new resolution to the original resolution in the temporal dimension. }

\added{To demonstrate the performance of this technique, here we \DIFdelbegin \DIFdel{infer }\DIFdelend \DIFaddbegin \DIFadd{test }\DIFaddend the original FNO, conv-FNO, and U-FNO at a temporal resolution that was not used in training. We generated 50 new data samples where each sample has 48-time steps; each time step is obtained through refining the original step size by 50\%. The original FNO is directly applied to the refined data set, while conv-FNO and U-FNO are modified as described above. Table 4 summarized the average $R^2$ score on the new data set in comparison to the original data set. }

\begin{table}[h]
\centering
\scriptsize
\caption{\added{$R^2$ score comparison for FNO, conv-FNO, and U-FNO models at the original and refined time steps. The scores are calculated by taking the average for 50 examples.}}
\begin{tabular}{c|ccc|ccc}
\hline
 & \multicolumn{3}{c|}{Gas saturation} & \multicolumn{3}{c}{Pressure buildup} \\
 & FNO       & conv-FNO     & U-FNO     & FNO       & conv-FNO      & U-FNO      \\ \hline
original time step & 0.989     & 0.991       & 0.993    &  0.995      &  0.995 &  0.996 \\
refined time step  & 0.986     & \textbf{0.988}       & 0.987    &  0.992      &  \textbf{0.993} &  0.992 \\
\hline
\end{tabular}\label{table:refined_time}
\end{table}

\added{While the performance for all models slightly decreases with the refined time steps, the FNO-based models still provide relatively good estimations at unseen times without additional training. Interestingly, we observe that conv-FNO performs the best for both gas saturation and pressure buildup. We hypothesize that this is because the Fourier layers in the conv-FNO are more efficient than in the original FNO due to the presence of the convolution layer. Meanwhile, the convolution layer is less influential to the outputs compared to the U-Net component in U-FNO, therefore provides the best results at refined time steps. }

\subsection{Fourier kernel visualization}

\begin{figure}[h]
    \centering
    \includegraphics[width=\textwidth]{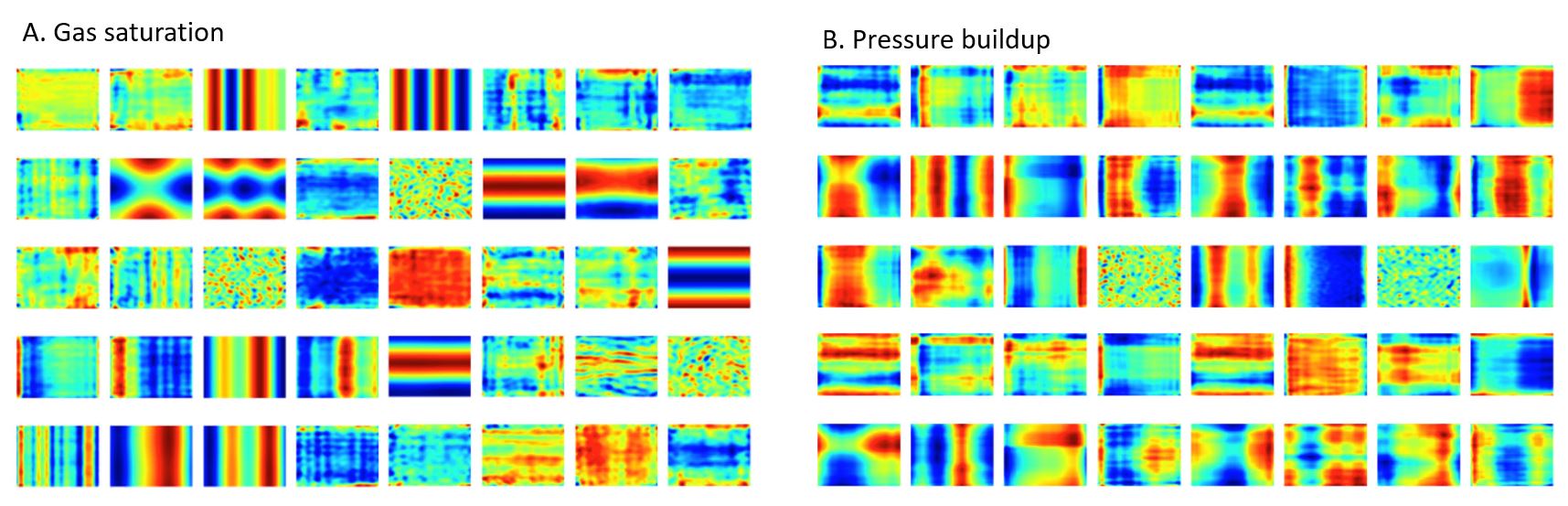}
    \caption{Visualizations of random selections of $(r,z)$ directional kernels  for trained A. gas saturation and B. pressure buildup models.}
    \label{fig:kernel}
\end{figure}

As described in Section 2, the Fourier path within each U-Fourier layer contains trainable kernel $R$ that is parameterized in the Fourier space. Here we provide visualizations for a random selection of the Fourier kernels in the trained gas saturation and pressure buildup models. Notice that unlike traditional CNN kernels that are generally small (\textit{e.g.}, $(3,3,3)$ or $(7,7,7)$), Fourier kernels are full field kernels that can be interpreted by any grid discretization. The kernels in this paper are 3D kernels with dimensions $(r,z,t)$ and the examples shown in Figure~\ref{fig:kernel} are the $(r,z)$ directional slices evaluated using the data discretization. Both gas saturation and pressure buildup models contain a wide variety of kernels from low to high frequency. We hypothesize that the asymmetry in the $r$ direction might be related to the gradually coarsening $r$-directional grid resolution, while the asymmetry in the $z$ direction might be related to the effects of buoyancy since CO$_2$ is less dense than water and tends to migrate to the top of the reservoir. 

\section{Conclusion}
This paper presents U-FNO, an enhanced Fourier neural operator for solving multiphase flow problems. We demonstrate that U-FNO predicts highly accurate flow outputs for a complex CO$_2$-water multiphase flow problem in the context of CO$_2$ geological storage. 

Through comparisons with the original FNO architecture~\citep{li2020fourier} and a state-of-the-art CNN benchmark~\citep{WEN2021104009}, we show that the newly proposed U-FNO architecture provides the best performance for both gas saturation and pressure buildup predictions. The U-FNO architecture enhances the training accuracy of a original FNO. At the same time, U-FNO maintains the excellent generalizability of the original FNO architecture. 
For the CO$_2$-water multiphase flow application described here, our goal is to optimize for the accuracy of gas saturation and pressure fields, for which the U-FNO provides the highest performance. 

The trained U-FNO model generates gas saturation and pressure buildup predictions that are \replaced{$6\times10^4$ times}{$10^5$ orders of magnitude} faster than a traditional numerical solver. The significant improvement in the computational efficiency can support many engineering tasks that require\deleted{s} repetitive forward numerical simulations. For example, the trained U-FNO model can serve as an alternative to full physics numerical simulators in probabilistic assessment, inversion, and site selection, tasks that were prohibitively expensive with desirable grid resolution using numerical simulation.

\section*{Code and data availability}
The python code for U-FNO model architecture and the data set used in training is available at \url{https://github.com/gegewen/ufno}. Web application \url{https://ccsnet.ai} hosts the trained U-FNO models to provide real time predictions.

\section*{Acknowledgments}
G. Wen and S. M. Benson gratefully acknowledges the supported by ExxonMobil through the Strategic Energy Alliance at Stanford University and the Stanford Center for Carbon Storage. Z. Li gratefully acknowledges the financial support from the Kortschak Scholars Program.
A. Anandkumar is supported in part by Bren endowed chair, LwLL grants, Beyond Limits, Raytheon, Microsoft, Google, Adobe faculty fellowships, and DE Logi grant. \added{The authors would like to acknowledge the reviewers and editors for the constructive comments}.

\appendix
\newpage
\section{Table of notations}\label{apx:table}
\begin{table}[h]
\centering\scriptsize
\caption{Table of notations.}
\begin{tabular}{|l|ll|}
\hline
                   & Notation & Meaning \\
\hline
Operator learning  & $D\in \mathds{R}^d$  & The spatial domain for the problem \\
                   & $a\in \mathcal{A}={D;\mathds{R}^{d_a}}$ & Input coefficient functions        \\
                   & $z\in \mathcal{Z}={D;\mathds{R}^{d_z}}$ & Target solution functions        \\
                   & $\mathcal{G}^\dag:\mathcal{A}\to \mathcal{Z}$   & The operator mapping from coefficients to solutions        \\
                   & n   & The size of the discretization        \\
                   & x   & Points in the spatial domain        \\
                   & $D_j=\{x_1,...,x_n\}\subset D$ & The discretization of $(a_j,u_j)$ \\
                   & $\mathcal{G}_\theta$   & An approximation of $\mathcal{G}^\dag$        \\
                   & $\mu$                  & A probability measure where $a_j$ is sampled from \\
                   & $C$                  & Cost function        \\
\hline
U-FNO              & $a(x)$         & The discretized data input        \\
                   & $z(x)$         & The discretized data output         \\
                   & $v_{l_j}(x),j=0,...,L$      & High dimensional representation of $a(x)$ in Fourier layers\\
                   & $v_{m_k}(x),k=0,...,M-1$  & High dimensional representation of $a(x)$ in U-Fourier layers \\
                   & $Q(\cdot)$  &  The lifting neural network\\
                   & $P(\cdot)$  &  The projection neural network \\
\hline
U-Fourier layer    & $\mathcal{K}$ & The Kernel integral operator applied on $v_l$ and $v_m$\\
                   & $R$ & The linear transformation applied on the lower Fourier modes\\
                   & $W$ & The linear transformation (bias term) applied on the spatial domain \\
                   & $U$ & The U-Net operator applied on $v_l$ and $v_m$\\
                   & $\sigma$         & The activation function \\
                   & $\mathcal{F},\mathcal{F}^{-1}$& Fourier transformation and its inverse\\
                   & $\kappa$& The kernel function learned from data \\
                   & $k_{max}$ & The maximum number of modes \\
                   & $c$ & The number of channels \\
\hline
Governing equation & $\eta=CO_2,water$   & Components of CO$_2$ and water         \\
                   & $p=w,n$             & Phases of wetting and non-wetting        \\
                   & $\varphi$              & The pore volume        \\
                   & $t$                 & Time        \\
                   & $S_p$               & The saturation of phase $p$        \\
                   & $\rho_p$            & The density of phase $p$         \\
                   & $X_p$               & The mass fraction of phase $p$         \\
                   & $\mathbf{F}$        & Flux        \\
                   & $q$                 & The source term        \\
                   & $P_p$               & The pressure of phase $p$           \\
                   & $k$                 & The absolute permeability        \\
                   & $k_{r,p}$           & The relative permeability of phase $p$        \\
                   & $\mu_p$             & The viscosity of phase $p$        \\
                   & $\mathbf{g}$        & Gravitational acceleration       \\
\hline
Sampling variable  & \multicolumn{2}{c|}{refer to Table 1} \\
\hline
\end{tabular}
\end{table}

\newpage
\section{Grid discretization}\label{apx:grid}
\begin{table}[h]
\centering\footnotesize
\caption{Vertical, radial, and temporal grid discretization for ECLIPSE numerical simulation runs. The radial grid width gradually coarsens as $d_{r_{min}} \times a_r^{j-1}$, for $j \in [1,...,i_r]$. The temporal step size gradually coarsens as $d_{t_{min}} \times a_t^{j-1}$, for $j \in [1,...,i_t]$.}
\begin{tabular}{|l|llll|}
\hline
Dimension    & Parameter             & Notation      & Value     & Unit  \\
\hline
Vertical ($z$) & box boundary        &  $z_{max}$    & 12\DIFaddbegin \DIFadd{.5 }\DIFaddend to 200    & m     \\
             & grid count            &  $i_z$        & 6 to 96      & -     \\
             & grid thickness        &  $d_z$        & 2.08      & m     \\
\hline
Radial ($r$) & box boundary        &  $r_{max}$      & 1,000,000 & m     \\
             & grid count            &  $i_r$          & 200       & -     \\
             & minimum grid width    &  $d_{r_{min}}$        & 3.6       & m     \\
             & amplification factor  &  $a_r$          & 1.035012  & -     \\
             & well radius & $r_{well}$ & 0.1 & m \\
\hline
Temporal ($t$) & total length        &  $t_{max}$    & 30        & years \\
             & step count            &  $i_t$        & 24        & -     \\
             & minimum step          &  $d_{t_{min}}$        & 1         & day   \\
             & amplification factor  &  $a_t$        & 1.421245  & -     \\
\hline
\end{tabular}
\end{table}

\newpage
\section{Heterogeneous permeability map statistical parameters and visualizations}\label{apx:perm}
\begin{figure}[h]
    \centering
    \includegraphics[width=\textwidth]{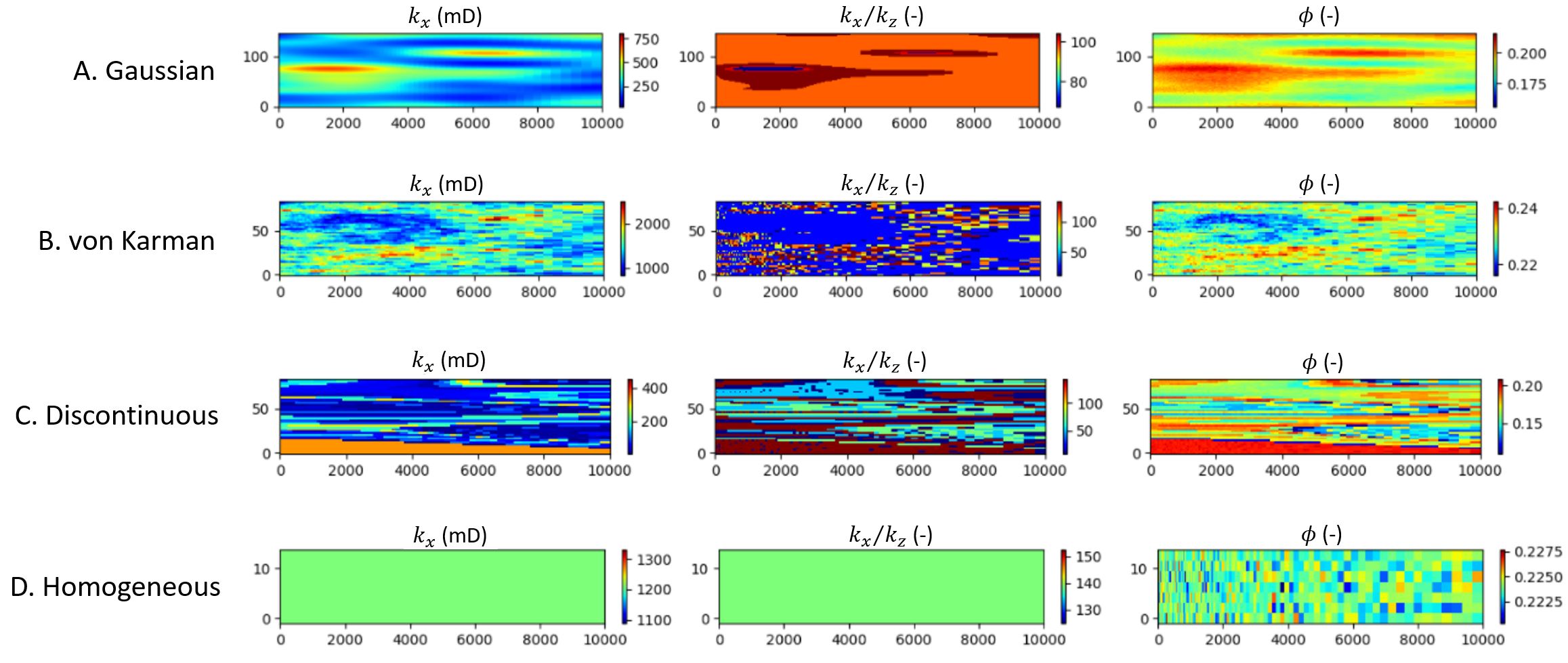}
    \caption{Horizontal permeability map, anisotropy map, and porosity map for A. Gaussian, B. von Karman, C. Discontinuous, and D. Homogeneous medium appearances.}
\label{fig:perm}
\end{figure}

\begin{table}[h]
\centering
\scriptsize
  \caption{Statistical parameters of horizontal permeability ($k_x$) maps generated by Stanford Geostatistical Modeling Software (SGeMS)~\citep{sgems}. We defined the medium appearance, spatial correlation, mean, standard deviation, and contrast ratio ($k_{high}/k_{low}$) in each map to create a large variety of permeability maps.}
\begin{tabular}{|lllllll|}
    \hline
    Medium & Parameter & Mean & Std & Max & Min & Unit\\
    \hline
    A. Gaussian & Field average & 30.8 & 58.3 & 1053 & 0.3 & mD \\
             & Vertical correlation & 7.3 & 3.6 & 12.5 & 2.1 & m \\
             & Horizontal correlation & 2190 & 1432 & 6250 & 208 & m \\
             & Contrast ratio & 4.01$\times$ 10$^4$ & 2.19$\times$ 10$^5$ & 3.00$\times$ 10$^6$ & 1.01 & - \\
  \hline
    B. von Karman & Field average & 39.9 & 54.4 & 867.9 & 1.8 & mD\\
      \citep{carpentier2009conservation}       & Vertical correlation & 7.2 & 3.5 & 12.5 & 2.1 & m\\
             & Horizontal correlation & 2.15$\times$ 10$^4$ & 1.40$\times$ 10$^4$ & 6.23$\times$ 10$^4$ & 208 & m\\             
             & Contrast ratio & 2.66$\times$ 10$^4$ & 1.54$\times$ 10$^5$ &  2.12$\times$ 10$^6$ & 1.00 & - \\     
  \hline
    C. Discontinuous & Field average & 80.8 & 260.2 & 5281 & 2.0 & mD\\
             & Vertical correlation & 7.2 & 3.6 & 12.5 & 2.1 & m\\
             & Horizontal correlation & 2176 & 1429 & 6250 & 208 & m\\             
             & Contrast ratio & 2.17$\times$ 10$^4$ & 1.51$\times$ 10$^5$ & 2.68$\times$ 10$^6$ & 1.01 & - \\               
  \hline
    D. Homogeneous & Field permeability & 327.7 & 478.1 & 1216 & 4.0 & mD\\
  \hline
  \end{tabular}
\label{sgems}
\end{table}

\newpage
\section{Darcy flow comparison}\label{apx:darcy}
\begin{figure}[h]
    \centering
    \includegraphics[width=0.6\textwidth]{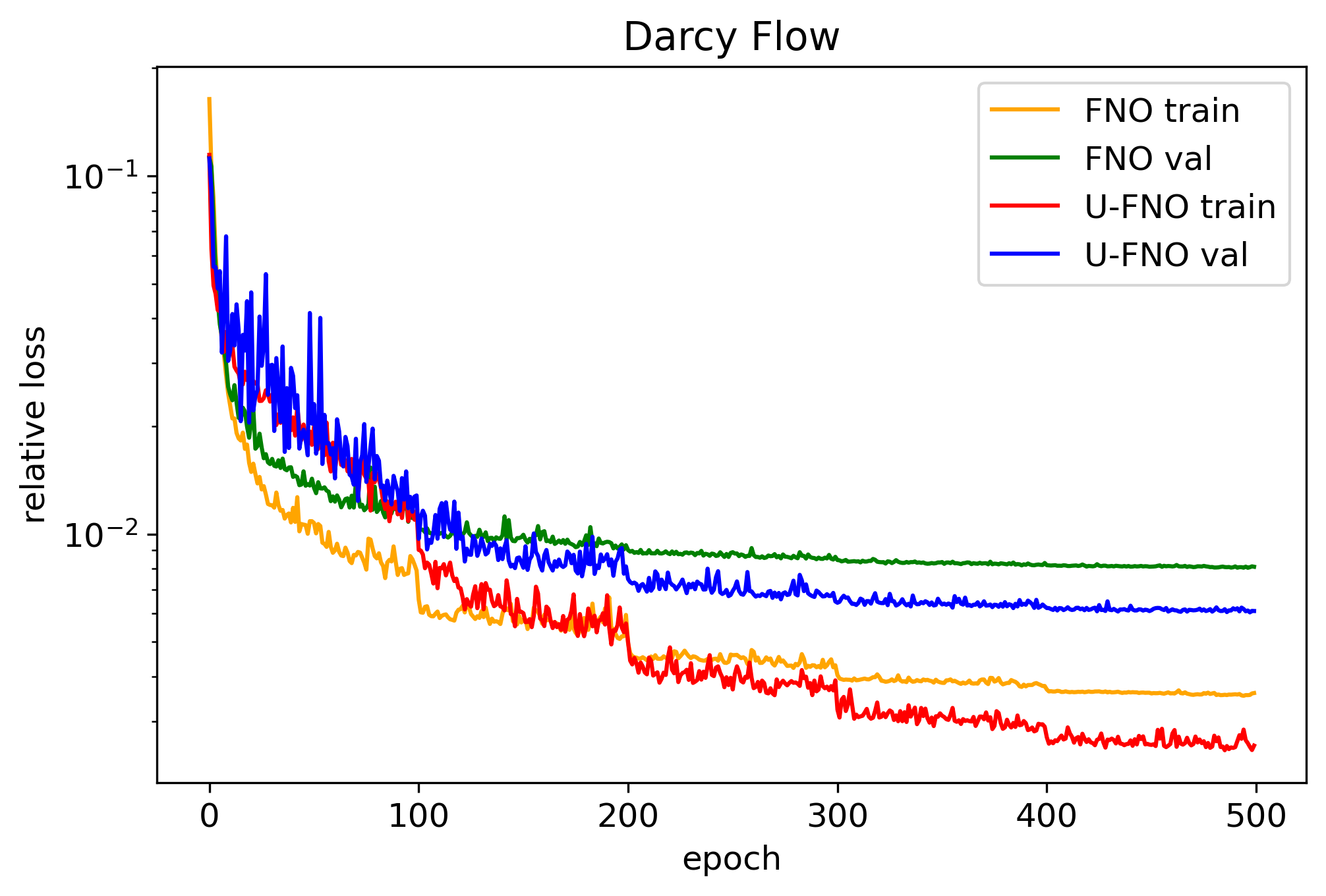}
    \caption{\added{Relative loss evolution vs. epoch for Darcy's flow problem by the FNO and U-FNO architecture.}}
    \label{fig:darcy}
\end{figure}

\added{Here we compared the performance of U-FNO with the original FNO on a steady-state Darcy's flow problem. Since this is a 2D problem, we used the 2D variation of U-FNO where we append a 2D U-Net to the 2D Fourier layer. The steady-state Darcy flow problem data set is provided in~\citep{li2020fourier}. Figure~\ref{fig:darcy} shows that U-FNO achieves lower relative loss than FNO for both training and validation set. We also compared the validation set relative loss with 4 state-of-the-art benchmark models in Table~\ref{tab:darcy}. The Darcy flow example demonstrates that the advantage of using U-FNO is not limited to multi-phase flow application.}

\begin{table}[h]
\centering
\scriptsize
\caption{\added{Relative loss in comparison to benchmark models. FCN is a Fully Convolution Network proposed in Zhu \& Zabaras, 2018~\citep{zhu2018bayesian}; PCANN is an operator method using PCA as autoencoder proposed in Bhattacharya et al, 2020~\citep{bhattacharya2020model}; GNO is the graph neural opeartor propsoed in Li et al., 2020~\citep{li2020neural}; and FNO is the original Fourier neural operator~\citep{li2020fourier}. The performance of these above models are listed in~\citep{li2020fourier}.} }
\begin{tabular}{c|c}
\hline
Model      & Validation set relative loss \\ \hline
FCN~\citep{zhu2018bayesian}   & 0.1097        \\
PCANN~\citep{bhattacharya2020model} & 0.0299        \\
GNO~\citep{li2020neural}   & 0.0369        \\
FNO~\citep{li2020fourier} & 0.0098        \\
\textbf{U-FNO (this paper)} & \textbf{0.0061}       \\ \hline
\end{tabular}\label{tab:darcy}
\end{table}

\newpage
\section{CNN benchmark model architecture}\label{apx:CNN}
\begin{table}[h]
  \centering
  \footnotesize
  \caption{CNN architecture. \texttt{Conv3D} denotes a 3D convolutional layer; \texttt{BN} denotes a batch normalization layer; \texttt{ReLu} denotes a rectified linear layer; \texttt{Add} denotes an addition with the identity; \texttt{UnSampling} denotes an unSampling layer that expands the matrix dimension using nearest neighbor method, and \texttt{Padding} denotes a padding layer using the reflection padding technique. In this model, the number of total parameters is 33,316,481 with 33,305,857 trainable parameters and 10,624 non-trainable parameters. To ensure a fair comparison with the FNO-based models, we performed hyper-parameter optimization on the CNN benchmark model and trained it with the same loss function (Equation~\ref{eq:loss}) as the FNO-based models.}
  \begin{tabular}{|lll|}
    \hline
    Part     & Layer     & Output Shape \\
    \hline
    Input    &  -                                  & (96,200,24,1) \\
    Encode 1 &  \texttt{Conv3D/BN/ReLu} & (48,100,12,32)  \\
    Encode 2 &  \texttt{Conv3D/BN/ReLu} & (48,100,12,64)  \\
    Encode 3 &  \texttt{Conv3D/BN/ReLu} & (24,50,6,128)  \\
    Encode 4 &  \texttt{Conv3D/BN/ReLu} & (24,50,6,128)  \\
    Encode 5 &  \texttt{Conv3D/BN/ReLu} & (12,25,3,256)  \\
    Encode 6 &  \texttt{Conv3D/BN/ReLu} & (12,25,3,256)  \\
   ResConv 1 &  \texttt{Conv3D/BN/Conv3D/BN/ReLu/Add} &  (12,25,3,256)  \\
   ResConv 2 &  \texttt{Conv3D/BN/Conv3D/BN/ReLu/Add} &  (12,25,3,256)  \\
   ResConv 3 &  \texttt{Conv3D/BN/Conv3D/BN/ReLu/Add} &  (12,25,3,256)  \\
   ResConv 4 &  \texttt{Conv3D/BN/Conv3D/BN/ReLu/Add} &  (12,25,3,256)  \\
   ResConv 5 &  \texttt{Conv3D/BN/Conv3D/BN/ReLu/Add} &  (12,25,3,256)  \\
    Decode 6 &  \texttt{UnSampling/Padding/Conv3D/BN/Relu} &  (12,25,3,256) \\
    Decode 5 &  \texttt{UnSampling/Padding/Conv3D/BN/Relu} & (24,50,6,256)  \\
    Decode 4 &  \texttt{UnSampling/Padding/Conv3D/BN/Relu} & (24,50,6,128)  \\
    Decode 3 &  \texttt{UnSampling/Padding/Conv3D/BN/Relu} & (48,100,12,128)  \\
    Decode 2 &  \texttt{UnSampling/Padding/Conv3D/BN/Relu} & (48,100,12,64)  \\
    Decode 1 &  \texttt{UnSampling/Padding/Conv3D/BN/Relu} & (96,200,24,32)\\
    Output   &  \texttt{Conv3D}                         & (96,200,24,1) \\
    \hline
  \end{tabular}
  \label{table:cnn}
\end{table}

\newpage
\section{FNO model architecture}\label{apx:FNO}
\begin{table}[h]
  \centering
  \caption{FNO model architecture. The \texttt{Padding} denotes a padding operator that accommodates the non-periodic boundaries; \texttt{Linear} denotes the linear transformation to lift the input to the high dimensional space, and the projection back to original space; \texttt{Fourier3d} denotes the 3D Fourier operator; \texttt{Conv1d} denotes the bias term; \texttt{Add} operation adds the outputs together; \texttt{ReLu} denotes a rectified linear layer. In this model, the number of total parameters is 31,117,541.}
  \footnotesize
  \begin{tabular}{|lll|}
    \hline
    Part     & Layer     & Output Shape \\
    \hline
    Input    & -                                    & (96,200,24,12) \\
    Padding &   \texttt{Padding}                    & (104, 208, 32, 12) \\
    Lifting &  \texttt{Linear}                      & (104, 208, 32, 36)  \\
    Fourier 1 & \texttt{Fourier3d/Conv1d/Add/ReLu}  & (104, 208, 32, 36)  \\
    Fourier 2 & \texttt{Fourier3d/Conv1d/Add/ReLu}  & (104, 208, 32, 36)  \\
    Fourier 3 & \texttt{Fourier3d/Conv1d/Add/ReLu}  & (104, 208, 32, 36)  \\
    Fourier 4 & \texttt{Fourier3d/Conv1d/Add/ReLu}  & (104, 208, 32, 36)  \\
    Fourier 5 & \texttt{Fourier3d/Conv1d/Add/ReLu}  & (104, 208, 32, 36)  \\
    Fourier 6 & \texttt{Fourier3d/Conv1d/Add/ReLu}  & (104, 208, 32, 36)  \\
    Projection 1   &  \texttt{Linear}                          & (104, 208, 32, 128) \\
    Projection 2   &  \texttt{Linear}                          & (104, 208, 32, 1) \\
    De-padding     &  -                                         & (96, 200, 24, 1) \\
    \hline
  \end{tabular}
  \label{table:apxfno}
\end{table}

\newpage
\section{Conv-FNO model architecture}\label{apx:convFNO}
\begin{table}[h]
  \centering
  \caption{Conv-FNO model architecture. The \texttt{Padding} denotes a padding operator that accommodates the non-periodic boundaries; \texttt{Linear} denotes the linear transformation to lift the input to the high dimensional space, and the projection back to original space; \texttt{Fourier3d} denotes the 3D Fourier operator; \texttt{Conv1d} denotes the bias term; \texttt{Conv3d} denotes a 3D convolutional operator; \texttt{Add} operation adds the outputs together; \texttt{ReLu} denotes a rectified linear layer. In this model, the number of total parameters is 31,222,625.}
  \footnotesize
  \begin{tabular}{|lll|}
    \hline
    Part     & Layer     & Output Shape \\
    \hline
    Input    & -                                  & (96,200,24,12) \\
    Padding &   \texttt{Padding}                                 & (104, 208, 32, 12) \\
    Lifting &  \texttt{Linear}                      & (104, 208, 32, 36)  \\
    Fourier 1 & \texttt{Fourier3d/Conv1d/Add/ReLu}  & (104, 208, 32, 36)  \\
    Fourier 2 & \texttt{Fourier3d/Conv1d/Add/ReLu}  & (104, 208, 32, 36)  \\
    Fourier 3 & \texttt{Fourier3d/Conv1d/Add/ReLu}  & (104, 208, 32, 36)  \\
    Conv-Fourier 1 & \texttt{Fourier3d/Conv1d/Conv3d/Add/ReLu}  & (104, 208, 32, 36)  \\
    Conv-Fourier 2 & \texttt{Fourier3d/Conv1d/Conv3d/Add/ReLu}  & (104, 208, 32, 36)  \\
    Conv-Fourier 3 & \texttt{Fourier3d/Conv1d/Conv3d/Add/ReLu}  & (104, 208, 32, 36)  \\
    Projection 1   &  \texttt{Linear}                          & (104, 208, 32, 128) \\
    Projection 2   &  \texttt{Linear}                          & (104, 208, 32, 1) \\
    De-padding     &  -                                         & (96, 200, 24, 1) \\
    \hline
  \end{tabular}
  \label{table:apxfnoconv}
\end{table}

\newpage
\section{U-FNO model architecture}\label{apx:UFNO}
\begin{table}[ht]
  \centering
  \footnotesize
  \caption{U-FNO model architecture. The \texttt{Padding} denotes a padding operator that accommodates the non-periodic boundaries; \texttt{Linear} denotes the linear transformation to lift the input to the high dimensional space, and the projection back to original space; \texttt{Fourier3d} denotes the 3D Fourier operator; \texttt{Conv1d} denotes the bias term; \texttt{UNet3d} denotes a two step 3D U-Net; \texttt{Add} operation adds the outputs together; \texttt{ReLu} denotes a rectified linear layer. In this model, the number of total parameters is 33,097,829.}
  \begin{tabular}{|lll|}
    \hline
    Part     & Layer     & Output Shape \\
    \hline
    Input    &  -                                 & (96,200,24,12) \\
    Padding &   \texttt{Padding}                  & (104, 208, 32, 12) \\
    Lifting &  \texttt{Linear}                      & (104, 208, 32, 36)  \\
    Fourier 1 & \texttt{Fourier3d/Conv1d/Add/ReLu}  & (104, 208, 32, 36)  \\
    Fourier 2 & \texttt{Fourier3d/Conv1d/Add/ReLu}  & (104, 208, 32, 36)  \\
    Fourier 3 & \texttt{Fourier3d/Conv1d/Add/ReLu}  & (104, 208, 32, 36)  \\
    U-Fourier 1 & \texttt{Fourier3d/Conv1d/UNet3d/Add/ReLu}  & (104, 208, 32, 36)  \\
    U-Fourier 2 & \texttt{Fourier3d/Conv1d/UNet3d/Add/ReLu}  & (104, 208, 32, 36)  \\
    U-Fourier 3 & \texttt{Fourier3d/Conv1d/UNet3d/Add/ReLu}  & (104, 208, 32, 36)  \\
    Projection 1   &  \texttt{Linear}                          & (104, 208, 32, 128) \\
    Projection 2   &  \texttt{Linear}                          & (104, 208, 32, 1) \\
    De-padding     &  -                                         & (96, 200, 24, 1) \\
    \hline
  \end{tabular}
  \label{table:apxfnounet}
\end{table}

\newpage
\section{Training, validation, and testing set accuracy}\label{Apx:scores}
\begin{table} [h]
\caption{\added{Training, validation, and testing data set performance summary. For each metric, $\mu$ denotes the average and $\sigma$ denotes the standard deviation. $MPE$ denotes the plume mean absolute error. $MRE$ denotes the field mean relative error as defined in~\citep{tang2020deep}. $R^2_{plume}$ denotes the $R^2$ score in the plume area.}}
\scriptsize
    \begin{subtable}[c]{\linewidth}
        \centering
        \caption{Gas saturation ($SG$)}
        \begin{tabular}{|c|cc|cccc|}
\hline
metric & data set      & value    & CNN & FNO & conv-FNO & \textbf{U-FNO} \\ \hline 
\multirow{6}{*}{$MPE$} 
& \multirow{2}{*}{train} & $\mu$       & 0.0200          & 0.0238          &  0.0191         & 0.0126      \\
&                        & $\sigma$    & 0.0110          & 0.0129          &  0.0101         & 0.0069      \\ 
& \multirow{2}{*}{val}   & $\mu$       & 0.0280          & 0.0265          &  0.0214         & 0.0154      \\
&                        & $\sigma$    & 0.0165          & 0.0153          &  0.0119         & 0.0097      \\ 
& \multirow{2}{*}{test}  & $\mu$       & \textbf{0.0299} & \textbf{0.0276} & \textbf{0.0224} & \textbf{\underline{0.0161}} \\
&                        & $\sigma$    & \textbf{0.0175} & \textbf{0.0160} & \textbf{0.0125} & \textbf{\underline{0.0105}} \\ 
\hline 
\multirow{6}{*}{$R^2_{plume}$}  
& \multirow{2}{*}{train} & $\mu$       & 0.982           &  0.971         &  0.980        & 0.989      \\
&                        & $\sigma$    & 0.019           &  0.029         &  0.021        & 0.013      \\ 
& \multirow{2}{*}{val}   & $\mu$       & 0.960           &  0.963         &  0.973        & 0.982      \\
&                        & $\sigma$    & 0.043           &  0.038         &  0.028        & 0.024     \\ 
& \multirow{2}{*}{test}  & $\mu$       & \textbf{0.955}  & \textbf{0.961} & \textbf{0.970} & \textbf{\underline{0.981}} \\
&                        & $\sigma$    & \textbf{0.047}  & \textbf{0.039} & \textbf{0.033} & \textbf{\underline{0.025}} \\ 
\hline 
        \end{tabular}
        \vspace{2mm}           
    \end{subtable}
    \begin{subtable}[c]{\linewidth}
        \centering
        \caption{Pressure buildup ($dP$)}
        \begin{tabular}{|c|cc|cccc|}
\hline
metric & data set   & value    & CNN & FNO & conv-FNO & U-FNO \\ \hline
\multirow{6}{*}{$MRE$} 
&  \multirow{2}{*}{train}& $\mu$       & 0.0064    & 0.0065   &  0.0067 &  0.0053     \\
&                        & $\sigma$    & 0.0060    & 0.0049   &  0.0055 &  0.0045     \\ 
& \multirow{2}{*}{val}   & $\mu$       & 0.0096    & 0.0083   &  0.0081 &  0.0072     \\
&                        & $\sigma$    & 0.0085    & 0.0063   &  0.0067 &  0.0058     \\ 
& \multirow{2}{*}{test}  & $\mu$       & \textbf{0.0089}    & \textbf{0.0082}   &  \textbf{0.0078} & \textbf{\underline{0.0068}}\\
&                        & $\sigma$    & \textbf{0.0063}    & \textbf{0.0052}   &  \textbf{0.0048} & \textbf{\underline{0.0045}}     \\ 
\hline
\multirow{6}{*}{$R^2$}  
&  \multirow{2}{*}{train}& $\mu$       & 0.990   & 0.991    &  0.992 &  0.994    \\
&                        & $\sigma$    & 0.029   & 0.014    &  0.018 &  0.014    \\ 
& \multirow{2}{*}{val}   & $\mu$       & 0.988   & 0.990    &  0.991 &  0.993    \\
&                        & $\sigma$    & 0.023   & 0.017    &  0.018 &  0.015    \\ 
& \multirow{2}{*}{test}  & $\mu$       & \textbf{0.987}   & \textbf{0.989}    & \textbf{0.990}  &\textbf{\underline{0.992}}    \\
&                        & $\sigma$    & \textbf{0.023}   & \textbf{0.021}    & \textbf{0.020}  &\textbf{\underline{0.020}}    \\ 
\hline 
        \end{tabular}
    \end{subtable}\label{tab:r2}
\end{table}

\newpage
 \bibliographystyle{elsarticle-num} 
 \bibliography{cas-refs}




\end{document}